\documentclass[useAMS,usenatbib]{mn2e}
\usepackage{graphicx}
\usepackage{pdflscape}
\begin{document}

\title[Testing the white dwarf MRR]{Testing the white dwarf mass-radius relation and comparing optical and far-UV spectroscopic results with \textit{Gaia} DR2, \textit{HST} and \textit{FUSE}}

\author[S.R.G.Joyce et al.]{S. R. G. Joyce$^{1}$\thanks{E-mail:
srgj1@le.ac.uk (SRGJ); mab@le.ac.uk (MAB)}, M. A. Barstow$^{1}$\footnotemark[1], S. L. Casewell$^{1}$, 
 M. R. Burleigh$^{1}$,\newauthor J. B. Holberg$^{2}$, H. E. Bond$^{3}$ \\
$^{1}$Dept. of Physics \& Astronomy, University of Leicester, University Road, Leicester, LE1 7RH \\
$^{2}$University of Arizona, LPL,Tucson, AZ, USA \\
$^{3}$Pennsylvania State University, University Park, PA, USA }


\pagerange{\pageref{firstpage}--\pageref{lastpage}} \pubyear{2017}

\maketitle

\label{firstpage}

\begin{abstract}
Observational tests of the white dwarf mass-radius relationship have always been limited by the uncertainty in the available distance measurements. Most studies have focused on Balmer line spectroscopy because these spectra can be obtained from ground based observatories, while the Lyman lines are only accessible to space based UV telescopes. We present results using parallax data from \textit{Gaia} DR2 combined with space based spectroscopy from \textit{HST} and \textit{FUSE} covering the Balmer and Lyman lines. We find that our sample supports the theoretical relation, although there is at least one star which is shown to be inconsistent. Comparison of results between Balmer and Lyman line spectra shows they are in agreement when the latest broadening tables are used. We also assess the factors which contribute to the error in the mass-radius calculations and confirm the findings of other studies which show that the spread in results for targets where multiple spectra are available is larger than the statistical error. The uncertainty in the spectroscopically derived log $g$ parameter is now the main source of error rather than the parallax. Finally, we present new results for the radius and spectroscopic mass of Sirius B which agree with the dynamical mass and mass-radius relation within 1$\sigma$.

\end{abstract}


\section{Introduction}

It has been almost 90 years since the theoretical mass-radius relationship (MRR) for white dwarfs was developed \citep{Chandrasekhar31}. This theory is fundamental to our understanding of white dwarfs (WDs) and is used in many studies where the masses and radii of the WD sample can not be measured independently. Despite its successful application to many areas of astrophysics, observational confirmation of the MRR itself has remained a great challenge. 

The MRR and WDs in general are useful for exploring many areas of astrophysics beyond the study of WDs themselves. Measuring the mass for large numbers of WDs allows us to investigate the WD mass distribution which contains information about the history of star formation in the galaxy. The observed WD mass distribution is sharply peaked at 0.6 M$_{\odot}$ (e.g., \citealt{Koester79, Bergeron_eta92, Tremblay_eta17}) rather than following a power law distribution  similar to the main sequence stars from which they form \citep{Salpeter55}. This implies that more massive stars return larger amounts of material to the interstellar medium when they end nuclear burning. Further evidence for this comes from studies of the initial-final mass relation (e.g., \citealt{Kalirai13}) which also rely heavily on the theoretical MRR to obtain the masses of the WDs.

Observational tests are required not only to confirm the validity of the MRR but also to test the refinements to the theory which take in to account the temperature and core composition. The thickness of the surface layer, which is composed of hydrogen for the DA white dwarfs in this study, must also be considered when calculating the ratio between mass and radius.

The original Chandrasekhar models were modified by \citet{HamadaSalpeter61} to include refinements to the internal equation of state and differing core compositions. More recently \citet{Wood95} and \citet{Fontaine_eta01} have considered the realistic thermal evolution of non-degenerate H and He envelopes as a function of photospheric temperature.

Increased temperatures and H-layer thickness produce models which predict systematically larger radii than the zero temperature models for a given mass. 
The differences in the predictions of the finite temperature models based on a range of temperature and H-layer thickness values has generally been smaller than the observational uncertainty, making it difficult to test the validity of these aspects of the models at the few per cent level.

Many previous studies have looked for observational constraints on the MRR using a range of methods. The spectroscopic method based on fitting the hydrogen lines was developed by \cite{Holberg85} and \cite{Bergeron_eta92}. It is the most widely applicable because it uses the prominent Balmer lines visible in the spectra of all DA white dwarfs to measure log $g$, and combines this with flux and distance measurements to calculate the radius and mass. The main source of uncertainty in these studies has always been the distance. The study by \cite{Schmidt96} found that the uncertainties in the mass-radius results obtained with pre-\textit{Hipparcos} parallaxes were too large to either confirm or disprove the MRR.

After 1996 \textit{Hipparcos} parallaxes became available and \citet{Vauclair97} and \cite{Provencal98} found that these considerably reduced the scatter in the observed MRR. Both studies concluded that the data generally agreed with the MRR, but the scatter was still too large to verify the differences between zero-temperature and evolutionary sequences. \cite{Provencal98} found that two stars in their sample were consistent with a thick H-layer, whereas \citep{Shipman_1997} found that 40 Eri B was only consistent with a thin H-layer. These were the first pieces of evidence indicating that DA WDs may have a range of hydrogen envelope thickness's rather than being all thick or all thin, which has implications for theories of WD formation.

A similar improvement came with the release of \textit{Gaia} DR1 \citep{Brown16} which included parallax measurements with a precision of $\sim0.5$ miliarcseconds (mas) compared to a few mas for \textit{Hipparcos}. \cite{Tremblay_eta17} used Balmer line spectroscopy with the \textit{Gaia} DR1 results and found that the observations were consistent with the theoretical MRR. 

\cite{Bedard17} used a larger sample of 219 stars with parallaxes gathered from \textit{Gaia} and ground based surveys. They found that the data supported the MRR, and the precision was good enough to identify significant outliers such as double-degenerate systems. 

Both \cite{Bedard17}  and \cite{Tremblay_eta17} have shown that with \textit{Gaia} DR1,  the data support the MRR, but for the majority of WDs, the precision achieved so far has not been enough to clearly distinguish between models with various temperatures or H-layer thickness due to the uncertainty in the spectroscopic parameters.  

\textit{Gaia} DR2 has now reduced parallax uncertainty to $\sim100$ microarcsec at 15 mag. However, \cite{Tremblay_eta17} have shown that even with the reduction in the parallax error, the remaining uncertainties in the spectroscopic parameters may still be too large to allow detailed tests of the MRR via the spectroscopic method.

Other methods of testing the MRR include the use of eclipsing binaries, the gravitational redshift method and the dynamical method. Excellent results have recently been achieved by using eclipsing binaries where the eclipses make it possible to measure very precise radii (e.g., \citealt{Parsons10, Bours16}). A recent study \citep{Parsons17} combined these radii with xshooter spectroscopy to measure the MRR for 26 WDs to a precision of 2.4 per cent (radius) and 2.7 per cent (mass). 

More high precision results have come from the dynamical method for the well known white dwarf Sirius B. \cite{Bond17} used 150 years of observations to determine the orbit of Sirius B and calculate its mass ($1.018 \pm 0.011$ M$_{\odot}$). This result is in agreement with the MRR and is particularly important because Sirius B is one of the few WDs that constrain the high mass end. A limitation of the dynamical method is that it does not provide an independent measurement of the radius as is the case with eclipsing binaries. The radius used for Sirius B in \cite{Bond17} was derived from \textit{HST} spectra and the detailed results will be presented in this paper.

Similar dynamical studies of Procyon B \citep{Bond15} and 40 Eri B \citep{Bond_40Eri_17, Mason17} also found that the mass derived from the binary orbits is in excellent agreement with the theoretical MRR. The precision achieved by this method is $\sim$ 1 per cent for Sirius B and Procyon B and 3 per cent for 40 Eri B.

The eclipsing and dynamical methods have provided the most precise tests of the MRR. Unfortunately, they are confined to studying WDs in near by and correctly oriented binaries and so are only applicable to limited samples. The spectroscopic method can be applied much more widely and easily.

To make progress with spectroscopic tests of the MRR, it will be necessary to reduce the uncertainty in the spectroscopic parameters beyond what has been achieved so far. One way of verifying results and identifying causes of uncertainty is to compare the spectroscopic mass for a particular WD to the values obtained from the other methods. 
Some of the best WDs for this kind of study are found in Sirius-Like Systems (SLSs) which consist of a WD and a main sequence star of spectral type K or earlier \citep{Holberg13}. Sirius B itself is particularly useful for studying the MRR because of its high mass and close proximity to Earth. Like other SLSs, the mass of Sirius B can in principal be derived using the spectroscopic, dynamical and gravitational redshift methods. Most WDs in SLSs are not included in \textit{Gaia} DR2 as individual stars. Fortunately, the main-sequence companions have parallax measurements and are at the same distance as the WD.

The small apparent binary separation (often less than 1 arc sec), and the faintness of the WD compared to the main-sequence companion in the optical, has made the study of many SLSs impossible until the advent of space based far-UV and optical spectroscopic instruments. Following a successful campaign to resolve the WD in more suspected SLSs \citep{Barstow_2001_resolving}, we were subsequently able to obtain the first optical spectra of the WD component in 3 new systems.

Even with \textit{HST}, many SLSs remain unresolved. Such WDs can only be studied in the far-UV where the contribution from the MS star is negligible. It is still possible to apply the spectroscopic method as the far-UV  covers the Lyman series of absorption lines. In principle, the spectroscopic method can be applied in the same way as to the Balmer lines in the optical. 

In this study we use Balmer/Lyman line fitting to test the MRR and also to cross check the results from optical and far-UV spectra. 
It is vital to test the validity of the Lyman line results because a large archive of WD spectra taken by \textit{FUSE} exists from which we can derive accurate log $g$ and temperatures (e.g., \citealt{Barstow_eta03, Barstow10}). The WDs in this archive cover a wide temperatures range (16,000 - 77,000 K) and will be ideally suited for testing the temperature dependence of the MRR. Until recently, the necessary  distance information for this sample was not available. Now that the parallax measurements needed to employ the \textit{FUSE} data for testing the MRR have arrived, it is important to compare the uncertainties and systematics of the hydrogen line fitting as applied to both Lyman and Balmer line analysis.

For Sirius B and HZ 43 B we have several spectra available which we use to assess the repeatability of spectroscopic results. This follows on from the recent work of \cite{Tremblay_eta17} which showed that for Wolf 485A the spread in $T_{\rm eff}$ and log $g$ values measured from several spectra was larger than the error estimates from fitting individual spectra. A follow-up paper will compare the results obtained here to measurements made using the gravitational redshift method in order to identify the strengths and limitations of each method.

\begin{table*}
\begin{minipage}{155mm}

\caption{List of white dwarfs and the spectra used in this study.}
\label{table:observations}
\begin{tabular}{ccccccc}

\hline
 & WD number & Name & HST obsID & FUSE obsID & Grating/slit & date\\
\hline
\textbf{HST + FUSE} &  &  &  &  &  & \\

\textbf{} & WD 0022$-$745 & \textbf{HD 2133 B} & obt802010 & - & G430L / 52x0.2$^{\prime\prime}$ & 2012-09-23\\
\textbf{} &  &  & obt802020 & - & G430L / 52x0.2$^{\prime\prime}$ & 2012-09-23\\
\textbf{} &  &  & - & B0550201000 &  & \\

\textbf{} & WD 0418$+$137 & \textbf{HR 1358 B} & obt808050 & - & G430L / 52x0.2$^{\prime\prime}$ & 2012-11-20\\
\textbf{} &  &  & obt808060 & - & G430L / 52x0.2$^{\prime\prime}$ & 2012-11-20\\

\textbf{} & WD 0512$+$326 & \textbf{14 Aur Cb} & otb804050 & - & G430L / 52x0.5$^{\prime\prime}$ & 2012-11-22\\
\textbf{} &  &  & obt804060 & - & G430L / 52x0.5$^{\prime\prime}$ & 2012-11-22\\
\textbf{} &  &  & - & A05407070 &  & \\

\textbf{} & WD 1314$+$293 & \textbf{HZ 43 B} & o57t01010 & - & G430L / 52x2$^{\prime\prime}$ & 1998-12-17\\
\textbf{} &  &  & o57t02010  & - & G430L / 52x2$^{\prime\prime}$ & 1998-12-19\\
\textbf{} &  &  & o69t07020 & - & G430L / 52x2$^{\prime\prime}$ & 2000-11-06\\
\textbf{} &  &  & o69t08020 & - & G430L / 52x2$^{\prime\prime}$ & 2000-12-10\\
\textbf{} &  &  & -  & M1010501000 &  & \\
\textbf{} &  &  & - & P1042301000 &  & \\
\textbf{} &  &  & - & P1042302000 &  & \\

\hline
\textbf{HST} &  &  &  &  &  & \\

\textbf{} & WD 0642$-$166 & \textbf{Sirius B} & obt801010 & - & G430L / 52x2$^{\prime\prime}$ & 2013-01-26\\
 &  &  & obt801020 & - & G430L / 52x2$^{\prime\prime}$ & 2013-01-26\\
 &  &  & obt801030 & - & G430L / 52x2$^{\prime\prime}$ & 2013-01-26\\
 &  &  & obt801040 & - & G430L / 52x2$^{\prime\prime}$ & 2013-01-26\\

\hline
\textbf{FUSE} &  &  &  &  &  & \\
\textbf{} & WD 0226$-$615 & \textbf{HD 15638} & - & A05402010 &  & \\
\textbf{} & WD 0232$+$035 & \textbf{Feige 24} & - & P10405040 &  & \\

\textbf{} & WD 0353$+$284 & \textbf{RE 0357} & - & B05510010 &  & \\
\textbf{} & WD 1021$+$266 & \textbf{RE 1024} & - & B05508010 &  & \\
\textbf{} & WD 1921$-$566 & \textbf{REJ 1925} & - & A05411110 &  & \\
\textbf{} & WD 2350$-$706 & \textbf{HD 223816} & - & A05408090 &  & \\

\hline
\end{tabular}
\end{minipage}
\end{table*}


\begin{table*}
\begin{minipage}{140mm}

\caption{Comparison of parallax values from \textit{Hipparcos} (new reduction, \citealt{vanLeeuwen07}) and \textit{Gaia} DR2. Binary parameters are from \citep{Holberg13} and references therein.}
\label{table:parallax}
\begin{tabular}{cccccc}

\hline
 Name & Hipparcos $\pi$  & Gaia DR2 $\pi$   & Orbital period & Main sequence SpT & Apparent separation ($\rho$)\\
 &(mas)&(mas) & (years) & & (arc sec) \\
\hline

\textbf{HD 2133 } & 7.32 $ \pm $ 0.93 &  7.641 $ \pm $ 0.027 & 665.03 & F7V & 0.6\\

\textbf{HD 15638} & 4.85 $ \pm $ 0.78 & 6.426 $ \pm $ 0.178 & $<$52.3 & F6V & $<$0.08\\
\textbf{Feige 24} & 10.9 $ \pm $ 3.94 & 12.669 $ \pm $ 0.054 & - & dM & -\\

\textbf{RE 0357} & n/a  & 9.287 $ \pm $ 0.076 & - & K2V & Unresolved\\
\textbf{HR 1358 } & 21.09 $ \pm $ 0.51 & 21.052 $ \pm $ 0.077 & 274.53 & F6V & 1.276\\

\textbf{14 Aur C} & 9.63 $ \pm $ 2.92 & 12.246 $ \pm $ 0.093 & 2432.72 & F2V & 2.0\\

\textbf{Sirius } & 379.21 $ \pm $ 1.58 & n/a & 50.1 & A0V & 7.5\\
\textbf{RE 1024} & n/a & 6.709 $ \pm $ 0.080 & - & F0V & $<$0.08\\
\textbf{HZ 43} & 25.96 $ \pm $ 6.38 & 16.756 $ \pm $ 0.074 & - & M3.5V & -\\
\textbf{REJ 1925} & n/a & 7.639 $ \pm $ 0.121 & 118.63 & G5V & 0.217\\
\textbf{HD 223816} & n/a  & 6.586 $ \pm $ 0.030 & - & G0V & 0.574\\

\hline
\end{tabular}
\end{minipage}
\end{table*}


\section[]{Observations}
\subsection{Overview}

The majority of the WDs in this sample were not included in the samples of \cite{Tremblay_eta17} and \cite{Bedard17} because optical spectroscopy was not available. The exceptions to this are Sirius B, Feige 24 and HZ 43 B. The optical spectra for Sirius B and HZ 43 B used in this analysis are from \textit{HST} and provide a comparison to the ground based results presented in those studies. Optical spectra for 14 Aur Cb, HD 2133 B and HR 1358 B were obtained by \textit{HST} and no previous optical spectra exist. 

The sample contains 11 targets, 3 of which have both \textit{HST} and \textit{FUSE} spectra available. This subset of targets will be used to study potential systematic differences between the results from the Lyman and Balmer lines. Sirius B and HZ 43 B both have 4 individual \textit{HST} spectra available taken with the G430L grating.  All of these targets are DA white dwarfs with hydrogen dominated atmospheres. 

We make use of spectra listed in Table \ref{table:observations} from \textit{HST} and \textit{FUSE}  as well as \textit{Gaia} DR2 parallax data in Table \ref{table:parallax}. The \textit{HST} data consist of STIS spectra taken with the G430L grating covering the 2900$-$5700  \AA\  range which includes the Balmer lines from $\beta$ to the series limit. The \textit{FUSE} spectra cover the wavelength range 912$-$1180 \AA\ which corresponds to the Lyman series from Lyman $\beta$ to the series limit.

\subsection{Spectra}

\textit{HST} data exists for 5 targets which were all observed with the G430L grating as part of program 12606 in cycle 19 (PI Barstow), except for HZ 43 B which was observed as part of the calibration of the \textit{HST} flux standards \citep{Bohlin95, Bohlin14} for program 8066 and 8849. Sirius B and HZ 43 B have exceptionally high quality spectra with 4 spectra each at  S/N $> 100$ due to their brightness. The number and quality of the spectra for these 2 targets allow us to test the intrinsic reproducibility of measurements from repeated observations of the same target.

There are 6 targets for which only \textit{FUSE} spectra are available. Processing of the data followed the procedures in \citet{Barstow_eta02, Barstow_eta03}. The \textit{FUSE} spectra have been re-binned to a resolution of 0.04 \AA\ because the spectra provided by the MAST archive are oversampled.

\subsection{Parallax}

The distances used in this study are calculated from the parallax measurements shown in Table \ref{table:parallax} provided by the \textit{Gaia} satellite (Prusti et al 2016). The \textit{Gaia} mission measures the positions of over a billion stars by repeatedly scanning the whole sky. The changes in position of each star over the course of the 5 year mission will allow the parallax and proper motion to be measured to an accuracy of a few micro-arc seconds compared to the mili-arc second accuracy achieved by the \textit{Hipparcos} mission. 

\textit{Gaia} data release 2 (DR2) was made publicly available on April 25th 2018 \citep{Gaia_DR2}. It is an improvement on the preliminary data release (DR1) which had to rely on measurements from the \textit{Hipparcos} and Tycho catalogues to provide a long enough baseline \citep{Brown16} and only included the first 14 months of \textit{Gaia} data. DR2 benefits from 22 months of \textit{Gaia} observations and improved calibration.  

The WDs in this sample are in close binaries and none except HZ 43 B are included in the \textit{Gaia} catalogue as individual stars. For these Sirius-Like Systems we can use the parallax measurement for the MS star to calculate the distance for the WD since the difference in distance is negligible.

\section{Analysis}

\subsection{Fitting procedure}

Testing the MRR requires accurate measurement of the mass and radius for a sample of WDs. We make use of the spectroscopic method developed by \cite{Holberg85, Bergeron_eta92}. The spectroscopic method is based on measuring the depth and broadening of the hydrogen absorption lines observed in the spectra of DA WDs. The amount of broadening in the absorption lines is a direct consequence of the gravitational field of the WD which keeps the atmosphere at such high pressure that the energy levels of the hydrogen atoms are distorted, causing them to absorb photons of a wider range of frequencies. The depth and shape of the lines also depends on the temperature which determines what proportion of atoms in the atmosphere are in a particular excitation state.

By fitting models generated by stellar atmosphere codes such as \textsc{tlusty} we can determine the best fitting values for log $g$ and $T_{\rm eff}$ which will reproduce the observed line shapes. The model fitting also includes a normalisation parameter, which enables us to calculate the radius of the star if its distance is known.

Our method differs slightly from that of \cite{Bergeron_eta92} in that the spectra are not normalised to remove the continuum. Instead, the procedure developed in \citet{Marsh97} and applied by \citet{Barstow05} for \textit{HST} spectra of Sirius B and \cite{Barstow_eta03, Barstow10} for the \textit{FUSE} data is to remove sections of the spectrum which are between the absorption lines and therefore contain only data points from the continuum. The absorption lines are the most sensitive to the log $g$ and $T_{\rm eff}$ parameters so we restrict our fitting to only include these regions of the spectrum.

Model grids produced from the \textsc{tlusty} models are loaded in to the \textsc{xspec} \citep{Arnaud96} fitting software. The spectra are fitted following the standard $\chi^{2}$ minimisation procedure.

\subsubsection{Fitting \textit{HST} Balmer line spectra}

The spectra were checked for any contamination from the bright main sequence companion star. This was found to be negligible for all targets except HR 1358 B. Therefore, a special background subtraction method had to be applied to the spectra of this star \citep{Joyce17}.

Fitting results are shown in Table \ref{table:fitting_results}. The results for targets with multiple spectra are taken as the average from fitting each of the spectra individually. Where a target has only one spectrum the quoted errors are the statistical uncertainty in the fit as calculated by \textsc{xspec}. When multiple spectra have been averaged the quoted error is the error in the average.

\subsubsection{Fitting \textit{FUSE} Lyman line spectra}

\begin{figure}
\includegraphics[width=83mm]{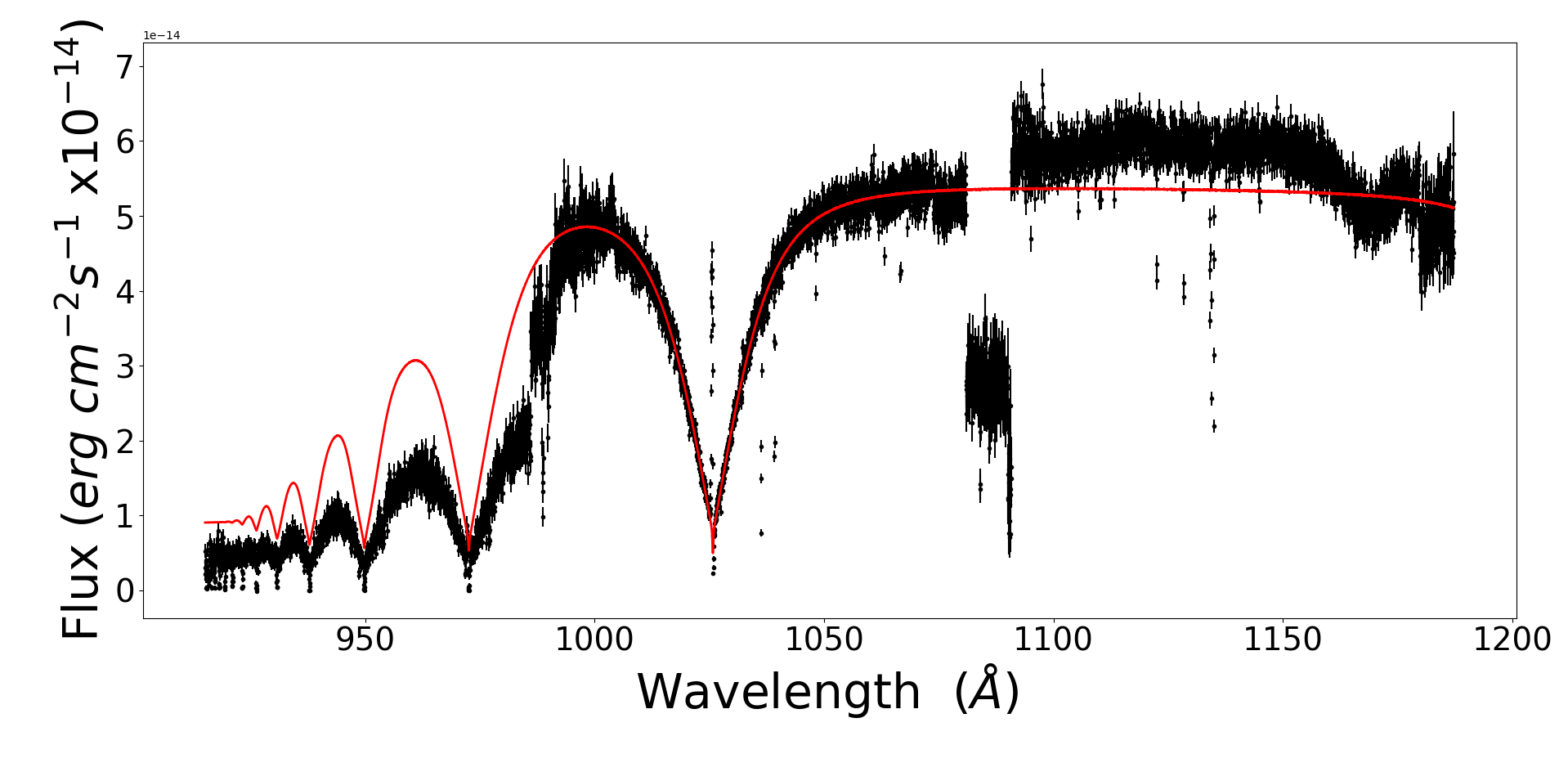}

\caption{\textit{FUSE} spectrum of 14 Aur Cb showing the difference in detected flux for sections of the spectrum recorded by different channels. The sections between 910-990 \AA\ and 1080-1100 \AA\ have a lower flux compared to the section between 990-1080 \AA\ (Lyman-$\beta$) due to the target not being aligned in those channels for the full duration of the exposure. The red line is a model normalised to the Lyman-$\beta$ line to highlight the difference in flux between this line and the rest of the Lyman series towards shorter wavelengths.}
\label{fig:14_Aur_spec}
\end{figure}

For the Lyman line spectra a slightly different fitting method has to be utilised as the satellite had different channels and detectors to record each part of the spectrum. Full details of the optical arrangement for \textit{FUSE} are given in \citet{Moos00} but some details of particular relevance are discussed here. 

There are 4 channels with different mirror and grating coatings which are optimized to reflect light in certain wavelength ranges. During operations it was found that it was not always possible to keep all 4 channels correctly pointed at the target for the whole exposure, leading to a loss of flux in some wavelength ranges. The consequence for this analysis is that the flux measured in a  spectrum can vary depending on which region of the spectrum is used. This is most noticeable in the spectrum of 14 Aur Cb shown in Fig. \ref{fig:14_Aur_spec} where the flux in region 980$-$1082 \AA\ is clearly lower than the flux in the rest of the spectrum. This spectrum is made up of several sections according to the wavelength ranges recorded by the different instrument channels. The model (red line) shows the normalisation when fitted to the Lyman-$\beta$ line only and is not a good fit to the rest of the Lyman lines which are recorded by a different channel.

Our model only has a single normalisation parameter and cannot account for the large difference in flux between different regions. The method developed to overcome this is to fit a separate model to each section of the spectrum which allows the normalisation to be altered for each  independently. The models are set up so that the $T_{\rm eff}$, log $g$ and abundance parameters are linked and cannot vary independently. Each channel/detector region of the spectrum is assigned to a separate model so that the model only fits a section with consistent flux. The other parameters in the models are the wavelength shift $z = \Delta\lambda / \lambda$ and the normalisation factor. The $z$ and normalisation parameters can vary independently so the resulting fit can adjust to the different flux levels but log $g$ and $T_{\rm eff}$ are still derived from fitting all of the available absorption lines simultaneously. The final normalisation value was selected from the sections of the spectrum which had the most flux, since flux could only be lost, not gained. For targets with both \textit{HST} and \textit{FUSE} data, the \textit{HST} value was used as it is more reliable.

Another issue that affects the Lyman lines but not the Balmer lines is geocoronal emission which causes strong narrow emission lines in the cores of the broad WD absorption lines (see Fig. \ref{fig:14_Aur_spec} at $\sim1025$ \AA). These emission lines are not included in our model since they are not emitted by the WD. Therefore, the emission lines are excluded when fitting the model to the absorption line. After exclusion of the spikes the spectra that are actually used for fitting have small gaps in the core of each Lyman absorption line. 

\subsection{White dwarf synthetic spectra}

For this study we use stellar models generated with the \textsc{tlusty} \citep{Hubeny95} code and the resulting spectral models are calculated using \textsc{synspec} \citep{HubenyLanz17}.

The model fitting results presented in this paper were obtained using a non-LTE pure-hydrogen WD grid. This grid covers the temperature range 18,000$-$80,000 K and log $g$ 7--9. This grid uses updated broadening tables \citep{Tremblay09} and includes additional updates from Tremblay in 2015 (private communication). All models were generated to cover both the Lyman and Balmer line regions covering the wavelength range 3000$-$7500 \AA\ so that the same model grid could be used for fitting both the \textit{HST} and \textit{FUSE} data and avoid possible systematic differences.

For the Balmer line spectra an initial fit was done with a coarse grid which has a lower resolution in log $g$ space of 0.25. This allowed the grid to cover the full range of possible $T_{\rm eff}$ and log $g$ values. Once an initial fit had been done, a high resolution grid with log $g$ spacing of 0.01 was produced for each target covering a smaller range of parameter space. 

For the Lyman line spectra there are existing values of $T_{\rm eff}$ and log $g$ available in 
\cite{Barstow_eta03, Barstow_eta14}. These results were obtained with a non-LTE H/He model grid using the older \cite{Lemke97} broadening tables. We repeated the fitting using an updated pure H grid based on the `Tremblay' broadening tables and the new $T_{\rm eff}$ and log $g$ results are presented in Table \ref{table:fitting_results}.

\subsection{Calculating mass and radius}
Calculating the radius of the WD requires measurements of the flux received and the distance from the WD to the observer. Our models give the predicted flux emitted per unit area by the WD and include a scaling factor to adjust for the fraction of flux per unit area at the distance of the detector. This scaling factor is listed in the normalisation column in Table \ref{table:fitting_results} and is used to calculate the radius using equation (\ref{eq:1}). 

\begin{equation}\label{eq:1}
normalisation = \frac{R^{2}}{D^{2}} 
\end{equation}

Spectra from \textit{HST} and \textit{FUSE} are flux calibrated so the normalization of the best fit model can be used to calculate the radius. 
The final ingredient for equation (\ref{eq:1}) is the distance $D$. This is provided by the \textit{Gaia} satellite using the parallax method. 

The radius is required as input to the mass calculation via equation (\ref{eq:2}).

\begin{equation}\label{eq:2}
g = \frac{GM}{R^{2}}
\end{equation}

Equation (\ref{eq:2}) is used to calculate the mass using the radius calculated in equation (\ref{eq:1}) combined with the log $g$ parameter found from model fitting of the spectrum. $G$ is the gravitational constant. 


\begin{table*}
\begin{minipage}{155mm}

\caption{Results of spectral fitting of the \textit{HST} and \textit{FUSE} spectra to determine the log $g$, $T_{\rm eff}$ and normalisation parameters. The normalisation column gives the scaling factor found from fitting the models which is related to the distance and radius of the WD via equation (\ref{eq:1}). These results are from fitting with a pure H non-LTE grid using the Tremblay broadening tables. \textit{FUSE} spectra have upper case obs ID while \textit{HST} spectra start with a lower case `o'. The error on the average values comes from the spread in results from the individual spectra divided by the square root of the number of spectra.}
\label{table:fitting_results}
\begin{tabular}{ccccc}

\hline
 obs ID & Name &  log $g$  & $T_{\rm eff}$  & Normalisation \\
 &     &  & (K)  & $(\frac{D^{2}}{R^{2}}) \times 10^{-21}$  \\
\hline

obt801010 & Sirius B  & 8.62  $\pm$  0.01 & 26102  $\pm$  63 & 0.463  $\pm$  0.002\\

obt801040 & Sirius B  & 8.61  $\pm$  0.01 & 25807  $\pm$  62 & 0.475  $\pm$  0.002\\

obt801030 & Sirius B  & 8.59  $\pm$  0.01 & 25885  $\pm$  64 & 0.473  $\pm$  0.002\\

obt801020 & Sirius B  & 8.57  $\pm$  0.01 & 25894  $\pm$  65 & 0.471  $\pm$  0.002\\

\textbf{Average} & \textbf{Sirius B} & \textbf{8.60}  $\pm$  \textbf{0.05} & \textbf{25922}  $\pm$  \textbf{296} & \textbf{0.471}  $\pm$  \textbf{0.013}\\
\\

M1010501000 & HZ 43 B & 7.921  $\pm$  0.006 & 50631  $\pm$  68 & 3.124e-3  $\pm$  7e-06\\

P1042301000 & HZ 43 B  & 7.897  $\pm$  0.004 & 50885  $\pm$  47 & 3.015e-3  $\pm$  5e-06\\

P1042302000 & HZ 43 B & 7.939  $\pm$  0.003 & 51110  $\pm$  33 & 2.952e-3  $\pm$  3e-06\\

\textbf{Average} & \textbf{HZ 43 B}  & \textbf{7.92}  $\pm$  \textbf{0.04} & \textbf{50875}  $\pm$  \textbf{414} & \textbf{3.03e-3}  $\pm$  \textbf{1.5e-4}\\
\\
o69u070 & HZ 43 B  & 7.90  $\pm$  0.03 & 51747  $\pm$  411 & 3.04e-3  $\pm$  2.5e-05\\

o69u080 & HZ 43 B  & 7.86  $\pm$  0.04 & 50943  $\pm$  387 & 3.06e-3  $\pm$  2.5e-05\\

o57t020 & HZ 43 B  & 7.88  $\pm$  0.05 & 50796  $\pm$  608 & 3.06e-3  $\pm$  4.0e-05\\

o57t010 & HZ 43 B  & 7.93  $\pm$  0.04 & 51414  $\pm$  528 & 3.03e-3  $\pm$  3.3e-05\\
\textbf{Average} & \textbf{HZ 43 B}  & \textbf{7.89}  $\pm$  \textbf{0.07} & \textbf{51225}  $\pm$  \textbf{950} & \textbf{3.05e-3}  $\pm$  \textbf{2.8e-05}\\
\\

A05407070 & 14 Aur Cb  & 7.93  $\pm$  0.02 & 42438  $\pm$  95 & 1.45e-3  $\pm$  6e-06\\
\\
obt804050 & 14 Aur Cb  & 7.87  $\pm$  0.10 & 45357 $\pm$  943 & 1.38e-3  $\pm$  3.4e-05\\

obt804060 & 14 Aur Cb  & 7.96  $\pm$  0.11 & 46291  $\pm$  1196 & 1.36e-3  $\pm$  4.1e-05\\
\textbf{Average} & \textbf{14 Aur Cb}  & \textbf{7.92}  $\pm$  \textbf{0.06} & \textbf{45824}  $\pm$  \textbf{660} & \textbf{1.37e-3}  $\pm$  \textbf{1.7e-05}\\

\\
\

B0550201000 & HD 2133 B & 7.6  $\pm$  0.1 & 28276  $\pm$  70 & 6.36e-4  $\pm$  1.5e-05\\
\\
obt802010 & HD 2133 B & 7.83  $\pm$  0.07 & 29612  $\pm$  266 & 6.01e-4  $\pm$  1.3e-05\\

obt802020 & HD 2133 B  & 7.64  $\pm$  0.08 & 29836  $\pm$  299 & 5.92e-4  $\pm$  1.4e-05\\
\textbf{Average} & \textbf{HD 2133 B}  & \textbf{7.73}  $\pm$  \textbf{0.13} & \textbf{29724}  $\pm$  \textbf{158} & \textbf{5.97e-4}  $\pm$  \textbf{6e-06}\\
\\

obt808050 & HR 1358 B  & 8.14  $\pm$  0.04 & 20922  $\pm$  190 & 3.37e-3  $\pm$  6.3e-05\\

obt808060 & HR 1358 B  & 8.10  $\pm$  0.08 & 20657  $\pm$  183 & 3.51e-3  $\pm$  6.4e-05\\

\textbf{Average} & \textbf{HR 1358 B}  & \textbf{8.12}  $\pm$  \textbf{0.03} & \textbf{20790}  $\pm$  \textbf{187} & \textbf{3.44e-3}  $\pm$  \textbf{9.7e-05}\\
\\

A05408090 & HD 223816  & 7.83  $\pm$  0.01 & 73999  $\pm$  267 & 6.49e-4  $\pm$  3e-06\\
\\

B05510010 & RE 0357  & 7.87  $\pm$  0.03 & 33927  $\pm$  66 & 8.84e-4  $\pm$  9e-06\\

\\

B05508010 & RE 1024  & 7.51  $\pm$  0.02 & 37274  $\pm$  37 & 1.09e-3  $\pm$  3.5e-05\\
\\

A05411110 & REJ 1925  & 7.80  $\pm$  0.03 & 49037  $\pm$  263 & 6.25e-4  $\pm$  7e-06\\
\\

P10405040 & Feige 24  & 7.64  $\pm$  0.01 & 62835  $\pm$  119 & 3.24e-3  $\pm$  8e-06\\
\\

A05402010 & HD 15638  & 7.66  $\pm$  0.02 & 50110  $\pm$  203 & 4.75e-4  $\pm$  5e-06\\

\hline

\end{tabular}
\end{minipage}
\end{table*}


\begin{table*}
\begin{minipage}{185mm}

\caption{Mass and radius values calculated using the spectroscopic parameters listed in Table \ref{table:fitting_results}.}
\label{table:mass_radius}
\begin{tabular}{cccccccc}

\hline
&& \textbf{HST} && \textbf{FUSE} &\\
Name & obs ID & Radius & Mass & Radius & Mass & Distance\\
&&(0.01 R$_{\odot}$)&(M$_{\odot}$)&(0.01 R$_{\odot}$)&(M$_{\odot}$)&(pc)\\

\hline

Sirius B & obt801010 & 0.796 $\pm$ 0.004 & 0.954 $\pm$ 0.029 & - & - & 2.639 $\pm$ 0.01\\
Sirius B & obt801020 & 0.804 $\pm$ 0.004 & 0.874 $\pm$ 0.026 & - & - & 2.639 $\pm$ 0.01\\
Sirius B & obt801030 & 0.805 $\pm$ 0.004 & 0.924 $\pm$ 0.028 & - & - & 2.639 $\pm$ 0.01\\
Sirius B & obt801040 & 0.807 $\pm$ 0.004 & 0.962 $\pm$ 0.028 & - & - & 2.639 $\pm$ 0.01\\

\textbf{Average} & HST & 0.802  $\pm$  0.011 & 0.927  $\pm$  0.107 & - & - & 2.637  $\pm$  0.011\\
&&&&&&\\

HZ43 & M1010501000 & - & - & 1.48 $\pm$ 0.007 & 0.665 $\pm$ 0.011 & 59.68 $\pm$ 0.262\\
HZ43 & P1042301000 & - & - & 1.454 $\pm$ 0.006 & 0.608 $\pm$ 0.008 & 59.68 $\pm$ 0.262\\
HZ43 & P1042302000 & - & - & 1.438 $\pm$ 0.006 & 0.656 $\pm$ 0.008 & 59.68 $\pm$ 0.262\\
\textbf{Average} & FUSE & - & - & 1.457 $\pm$ 0.036 & 0.643 $\pm$ 0.065 & 59.68 $\pm$ 0.262\\

&&&&&&\\
HZ43 & o69u070 & 1.459 $\pm$ 0.009 & 0.614 $\pm$ 0.048 & - & - & 59.68 $\pm$ 0.262\\
HZ43 & o69u080 & 1.464 $\pm$ 0.009 & 0.568 $\pm$ 0.048 & - & - & 59.68 $\pm$ 0.262\\
HZ43 & o57t020 & 1.464 $\pm$ 0.011 & 0.59 $\pm$ 0.067 & - & - & 59.68 $\pm$ 0.262\\
HZ43 & o57t010 & 1.458 $\pm$ 0.01 & 0.661 $\pm$ 0.066 & - & - & 59.68 $\pm$ 0.262\\
\textbf{Average} & HST & 1.461 $\pm$ 0.009 & 0.607 $\pm$ 0.106 & - & - & 59.68 $\pm$ 0.262\\
&&&&&&\\
14 Aur Cb & A05407070 & - & - & 1.378 $\pm$ 0.011 & 0.59 $\pm$ 0.028 & 81.656 $\pm$ 0.62\\
&&&&&&\\
14 Aur Cb & obt804050 & 1.346 $\pm$ 0.0193 & 0.492 $\pm$ 0.126 & - & - & 81.656 $\pm$ 0.622\\
14 Aur Cb & obt804060 & 1.334 $\pm$ 0.023 & 0.595 $\pm$ 0.178 & - & - & 81.656 $\pm$ 0.622\\
\textbf{Average} & HST & 1.34 $\pm$ 0.013 & 0.541 $\pm$ 0.086 & - & - & 81.656 $\pm$ 0.622\\
&&&&&&\\

HD2133 B & B0550201000 & - & - & 1.464 $\pm$ 0.018 & 0.277 $\pm$ 0.069 & 130.876 $\pm$ 0.462\\
&&&&&&\\
HD2133 B & obt802010 & 1.423 $\pm$ 0.016 & 0.495 $\pm$ 0.088 & - & - & 130.876 $\pm$ 0.462\\
HD2133 B & obt802020 & 1.413 $\pm$ 0.018 & 0.32 $\pm$ 0.068 & - & - & 130.876 $\pm$ 0.462\\
\textbf{Average} & HST & 1.418 $\pm$ 0.009 & 0.398 $\pm$ 0.138 & - & - & 130.876 $\pm$ 0.462\\
&&&&&&\\

HR1358 B & obt808050 & 1.223 $\pm$ 0.0123 & 0.748 $\pm$ 0.072 & - & - & 47.502 $\pm$ 0.173\\
HR1358 B & obt808060 & 1.248 $\pm$ 0.012 & 0.711 $\pm$ 0.137 & - & - & 47.502 $\pm$ 0.173\\
\textbf{Average} & HST & 1.235 $\pm$ 0.018 & 0.729 $\pm$ 0.053 & - & - & 47.502 $\pm$ 0.173\\
&&&&&&\\

RE 0357 & B05510010 & - & - & 1.42 $\pm$ 0.014 & 0.543 $\pm$ 0.039 & 107.68 $\pm$ 0.876\\

HD 223816 & A05408090 & - & - & 1.717 $\pm$ 0.009 & 0.723 $\pm$ 0.019 & 151.846 $\pm$ 0.689\\

RE 1024 & B05508010 & - & - & 2.183 $\pm$ 0.043 & 0.559 $\pm$ 0.03 & 149.065 $\pm$ 1.782\\

REJ 1925 & A05411110 & - & - & 1.452 $\pm$ 0.024 & 0.524 $\pm$ 0.04 & 130.907 $\pm$ 2.065\\

Feige 24 & P10405040 & - & - & 1.993 $\pm$ 0.009 & 0.633 $\pm$ 0.012 & 78.935 $\pm$ 0.335\\

HD15638 & A05402010 & - & - & 1.504 $\pm$ 0.042 & 0.376 $\pm$ 0.026 & 155.618 $\pm$ 4.303\\

\hline
\end{tabular}
\end{minipage}
\end{table*}

\section{Results}

\begin{figure*}
\includegraphics[width=175mm]{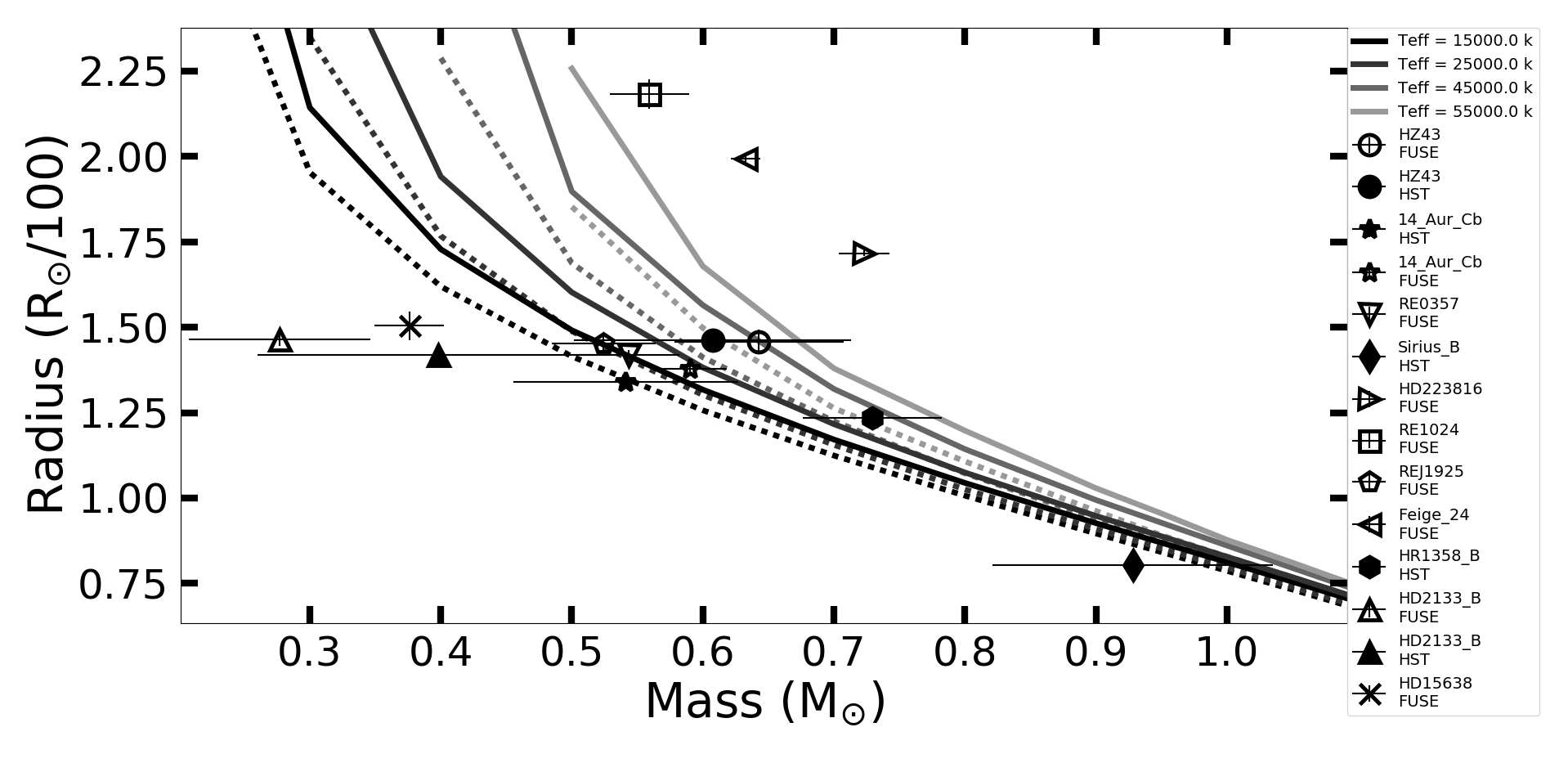}
 
\caption{The mass-radius relation with both \textit{HST} and \textit{FUSE} data. Data comes from Table \ref{table:mass_radius} and is based on fitting with the pure H non-LTE model grid using \citep{Tremblay09} broadening tables. Theoretical MRR models \citep{Fontaine_eta01} are shown for temperatures of 15,000, 25,000, 45,000 and 55,000 K from dark grey to light grey. Dashed lines are thin H-layer and solid lines are thick H-layer. Shapes of the symbols represent different white dwarfs as listed in the legend. The \textit{FUSE} Lyman line results are unfilled and the \textit{HST} Balmer line results are solid. Where a target has data available from both \textit{HST} and \textit{FUSE} they are plotted as 2 separate data points of matching shape.}
 \label{fig:MRR}
\end{figure*}

\begin{figure*}
\includegraphics[width=175mm]{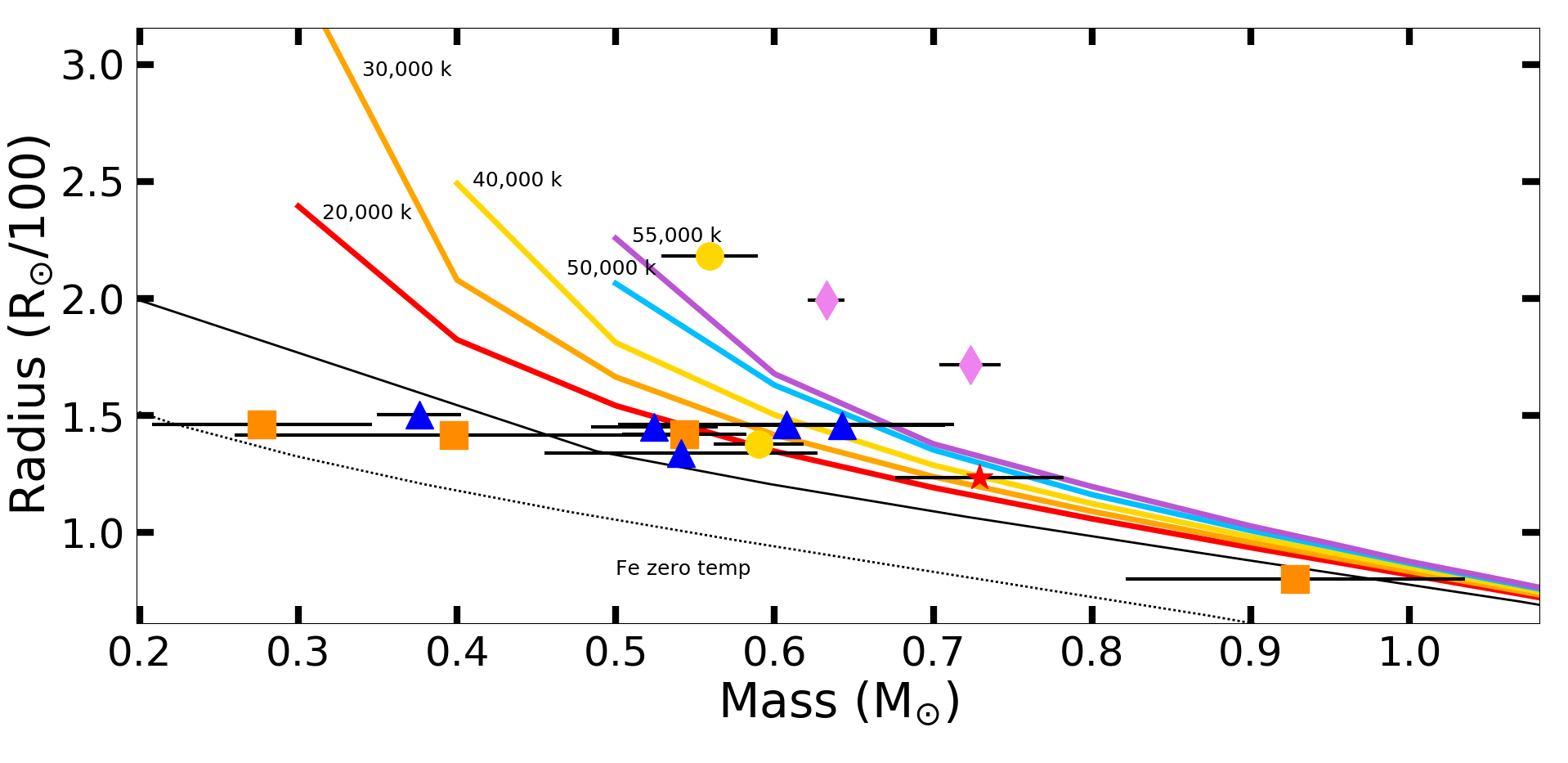}

\caption{The effect of temperature on the mass-radius relation. The mass-radius results are colour coded according to the temperature of the white dwarf. (Red star) 18,000 - 25,000 K, (Orange square) 25,000 - 35,000 K, (Yellow circle) 35,000-45,000 K ,(Blue triangle) 45,000-55,000 K, (Purple diamond) $>$ 55,000 K.
Two theoretical zero temperature mass-radius relations are shown as black lines. They are for core compositions of Fe (bottom) and carbon (top) \citep{HamadaSalpeter61}. Above the zero temperature relations are C/O core, thick H-layer models for temperatures of 20,000 - 55,000 K as indicated on the figure \citep{Fontaine_eta01}. The C/O core models are calculated for the temperatures in the middle of the ranges given for the data points and match the colour of the corresponding data points in the on-line version.}
\label{fig:MRR_Teff}
\end{figure*}

The results of fitting the spectra with the model grid are given in Table \ref{table:fitting_results}. For each target the results of fitting each individual spectrum are listed. The WDs in this sample cover a temperature range of 20,922 K for HR 1358 B to 73,999 K for HD 223816. Log $g$ ranges from 7.5 for RE 1024 to 8.6 for Sirius B.

The mass-radius results in Table \ref{table:mass_radius} were calculated by combining the atmospheric parameters in Table \ref{table:fitting_results} and the parallaxes in Table \ref{table:parallax}. The \textit{Gaia} DR2 parallax is used for all targets except Sirius which is too bright to be measured by \textit{Gaia}. The mass-radius results from Table \ref{table:mass_radius} are shown in Fig. \ref{fig:MRR}. The data are plotted with the theoretical MRR \citep{Fontaine_eta01} which was calculated for carbon/oxygen core WDs for a range of $T_{\rm eff}$. The temperatures plotted here are 15, 25, 45 and 55,000 K from dark to light grey. Dashed lines represent thin H-layer models $(q_{\rm H}=M_{\rm H}/M_{*}=10^{-10}$) and solid lines are thick H-layer models $(q_{\rm H}=M_{\rm H}/M_{*}=10^{-4}$). 

\textit{HST} Balmer line spectra are plotted as filled shapes and \textit{FUSE} Lyman line results are unfilled shapes. Each target is plotted as a different shape as shown in the legend. For targets that have Balmer and Lyman spectra they are plotted as two separate points of the same shape (filled or unfilled respectively).

To search for any effects due to temperature, we have plotted the mass-radius  results again in Fig. \ref{fig:MRR_Teff}. The WDs are binned into temperature ranges of 10,000 K as indicated by the marker shapes (and colours in the on-line version). The MRRs are for thick H-layer and are calculated for the temperature in the middle of each 10,000 K bin. The colours match the temperature ranges of the data points. We also plot the zero temperature carbon core and Fe core relations of \cite{HamadaSalpeter61} as the lowest (thin black) lines.


 \section{Discussion}


\subsection{Sources of uncertainty}\label{section:sources_of_uncertainty}

\begin{figure*}
\includegraphics[width=175mm]{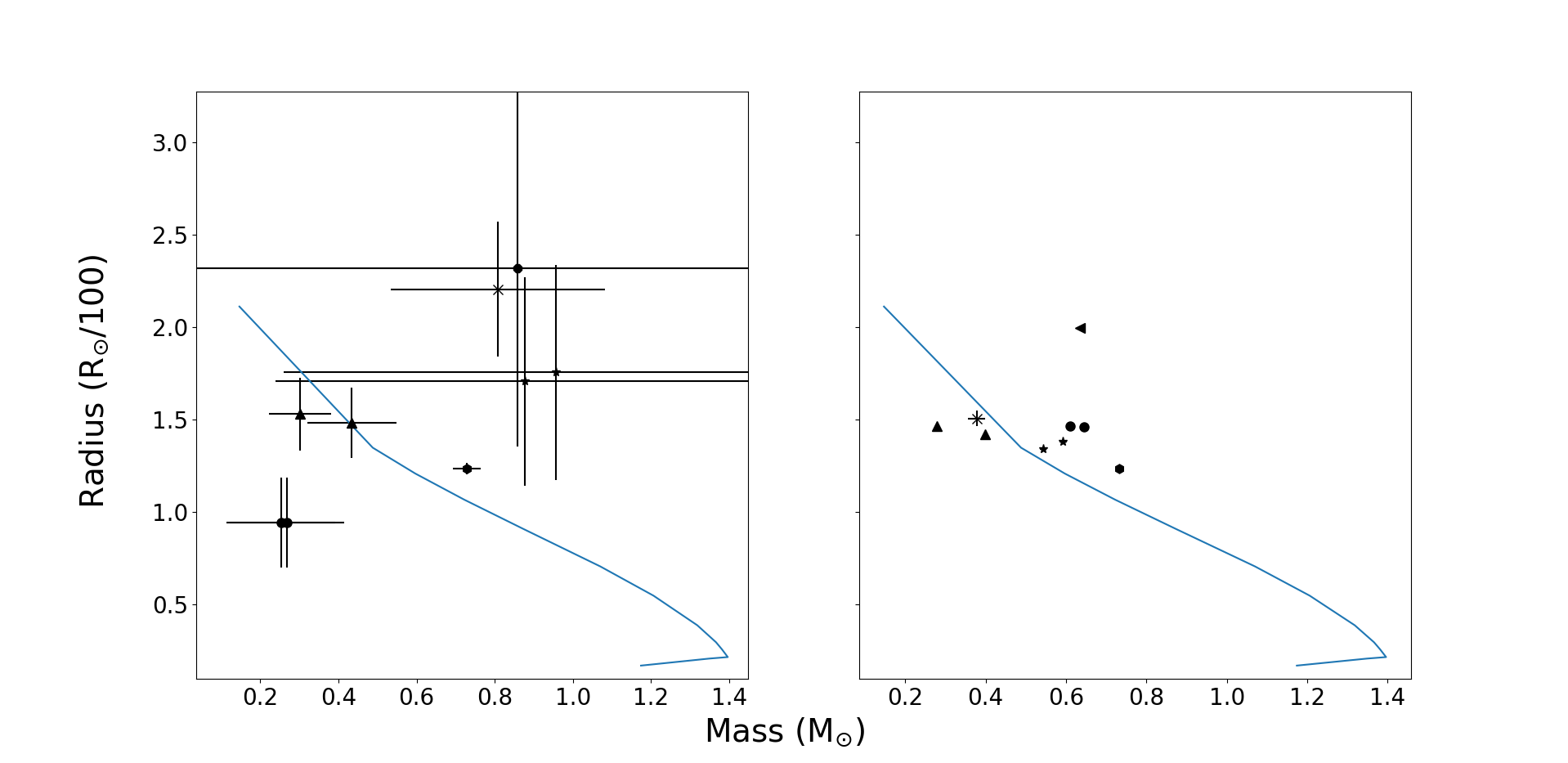}
 
 \caption{Comparison of results using parallax from \textit{Hipparcos} (left panel) and \textit{Gaia} DR2 (right panel). The blue line is the Hamada-Salpeter zero-temperature MRR for a C/O core WD. Error bars are calculated from the error in the parallax and do not include the error due to the spectroscopic fitting parameters. The DR2 errors are too small to be seen on this scale.}
  \label{fig:Hip_vs_gaia}
\end{figure*}

The currently available data is a vast improvement over the results that could be obtained using previously available parallaxes. However, there is clearly a great deal of uncertainty remaining. Here we investigate whether the main source of uncertainty is still the parallax and compare this error contribution to the error from the spectroscopic fitting parameters.

\subsubsection{Parallax}

As a comparison to show the effect of the improved parallax data, Fig. \ref{fig:Hip_vs_gaia} shows the MRR calculated using the \textit{Hipparcos} parallaxes (left panel) and the \textit{Gaia} DR2 parallaxes (right panel). Only the zero temperature \citep{HamadaSalpeter61} relation is plotted here for clarity. This plot only contains a subset of the WDs, which have parallaxes available in both catalogues. The errorbars are calculated from the uncertainty in the parallax alone, and do not include the error from the spectroscopic fitting. It clearly shows that the \textit{Gaia} parallaxes not only reduce the error bars to the point of being negligible, but also bring many of the targets into much closer agreement with the MRR. The distance measurements are no longer a major source of error. 
 

\subsubsection{Atmospheric parameters}

For the majority of targets, where only one spectrum is available, the uncertainty in the normalisation and log $g$ has been taken as the statistical error on the parameter found by the model fitting procedure. For the log $g$ parameter this gives an uncertainty of on average 0.03. However, this error range is smaller than the spread in log $g$ results found from fitting multiple spectra from the same target taken with the same instrument.
In Fig. \ref{fig:norm_spread} and Fig. \ref{fig:logg_spread} the top and bottom panels show the spread in $T_{\rm eff}$ and log $g$ from fitting multiple spectra of Sirius B and HZ 43 B. A similar test was carried out by \cite{Tremblay_eta17} for Wolf 485A (See their fig. 6). The spread in values for Wolf 485A found when using the same models and fitting technique is indicated by the black cross at the centre of each plot. We have set the axis of the plots to have the same width to aid direct comparison. The optical spectra for Sirius B (top panel) give a smaller spread than Wolf 485A. 
 The 4 Sirius B spectra have a log $g$ range of 0.046. The spectra for Sirius B are exceptionally high S/N ($\sim 200$) so it is possible that this spread in log $g$ results represents a lower limit for the log $g$ uncertainty using this method of spectral fitting. The log $g$ range for HZ 43 B is larger at $\sim$0.7 and is more indicative of the range that would apply to most WDs. In terms of standard deviation, the spread is 0.026$\sigma$ for HZ 43 B which is slightly smaller than the spread found from ground based spectra. \cite{Liebert_2005}  and \cite{Koester_2009} found the standard deviation in log $g$ from multiple spectra to be 0.038$\sigma$ and 0.09$\sigma$ respectively.
 

It is clear that the spread in values for HZ 43 B is the same as for Wolf 485 A in the optical. The far-UV on the other hand gives a much smaller spread. In Fig. \ref{fig:logg_spread}, (lower panel) it is particularly noticeable that the uncertainty associated with the individual FUSE results (stars) is much smaller than the HST results (circles). It was noted \citep{Barstow_eta03} that the Lyman lines gave smaller uncertainties in log $g$ and $T_{\rm eff}$ compared to the Balmer line spectroscopy. The Balmer line spectra in \cite{Barstow_eta03}  were ground based observations with S/N $\sim50-100$ similar to the \textit{HST} spectra used here. 
In Fig. \ref{fig:MRR_spread_Sir_HZ 43} the mass and radius have been calculated for each individual spectrum for Sirius B and HZ 43 B rather than taking the average. The top panel shows the spread in mass-radius results for Sirius B. The bottom panel compares the spread and uncertainty in the mass-radius values from the HZ 43 B Balmer line (filled circles) and Lyman line (empty circles) spectra. The spread in mass values derived from these two targets is $\sim$0.1 M$_{\odot}$. 
This shows that when the spread in parameter values taken from multiple spectra is used as an estimate of the measurement uncertainty, the resulting error range in the mass-radius calculations is still too large to be able to distinguish between theoretical MRRs with different core compositions or H-layer thickness using the spectroscopic method.


\subsubsection{Spectroscopic method precision and the MRR}

To evaluate the contribution of each of the input parameters as a percentage of the total error we have calculated the individual error contributions in Table \ref{table:error_percent}.

The distance and normalisation contribute to both the radius and mass calculations via equations (\ref{eq:1}) and (\ref{eq:2}) so their contribution to the radius error and the mass error are listed separately. The error contribution of each parameter was calculated by propagating the statistical uncertainty from the $\chi^{2}$ fitting through the mass-radius calculations. These are then converted to a percentage of the final error for ease of comparison. Our discussion of the spread in values from several spectra compared to the uncertainty in individual spectra has shown that the spread in values is much larger than the statistical uncertainty. In order to compare these two measures of uncertainty, Table \ref{table:error_percent} includes an `Average' row for each target. The error values in this row are based on the standard error in the mean for parameter values taken from several spectra of the same target. The `Average' uncertainty gives a more realistic estimate of the uncertainty involved with these measurements. It is important to compare this with the errors quoted for targets where only a single spectrum is available as the errors are likely to be underestimated for most targets.

For the radius calculations, the error due to the distance derived from the parallax is now less than or equal to the uncertainty from the normalisation. All targets except HZ 43 B have a distance error contribution of $\sim$50 per cent or less to the error in the average radius. For most targets the spread in the normalisation values is similar to the error in individual measurements so the percentage error contribution remains the same for the average radius. A notable exception is Sirius which has a parallax error of only 0.4 per cent due to its close proximity to Earth, so in this case the spread in normalisation is the dominant source of error at 78 per cent of the total error. 

For the mass calculations the log $g$ parameters is the dominant source of uncertainty contributing between 63-96 per cent of the total error when calculated from the spread in the average log $g$. The error in the distances is now the smallest source of uncertainty, contributing no more than 6 per cent to the average mass error. This confirms that \textit{Gaia} DR2 has reduced the distance errors to the point where they are no longer dominant, and supports the conclusion of \cite{Tremblay_eta17} that it is the atmospheric parameters which are now limiting the accuracy of the mass-radius measurements. For all targets in this sample, the results for radius measurements show greater consistency than the mass measurements regardless of whether Lyman or Balmer spectra are used. 


\begin{table*}
\begin{minipage}{165mm}

\caption{The per cent error contribution of each input parameter to the final error in the mass-radius calculations. The values in the main rows are calculated from the statistical error on each parameter from fitting individual spectra. The values in the 'Average' rows are based on the standard error in the mean parameter values found from fitting several spectra of the same target.}
\label{table:error_percent}
\begin{tabular}{cccccccc}

\hline
Name & Distance err  & Normalisation err  & Radius  & Mass  & Distance err  & Normalisation err  & g err \\

& $\%$ (R) & $\%$ (R) & (0.01 R$_{\odot}$) & (M$_{\odot}$) & $\%$ (M) & $\%$ (M) & $\%$ (M) \\
\hline

HZ 43 B 
 & 79 & 20 &  1.48$\pm$0.0067 & 0.67$\pm$0.01 & 31 & 7 & 60 \\
HZ 43 B
 & 85 & 14 &  1.45$\pm$0.0065 & 0.61$\pm$0.01 & 40 & 6 & 52 \\
HZ 43 B
 & 89 & 10 &  1.44$\pm$0.0064 & 0.66$\pm$0.01 & 48 & 5 & 45 \\
\textbf{Average} & 15 & 84 &  1.46$\pm$0.0364 & 0.64$\pm$0.06 & 5 & 30 & 63 \\
\hline
HZ 43 B
 & 51 & 48 &  1.46$\pm$0.0088 & 0.61$\pm$0.05 & 7 & 6 & 86 \\
HZ 43 B
 & 52 & 47 &  1.46$\pm$0.0087 & 0.57$\pm$0.05 & 6 & 5 & 87 \\
HZ 43 B
 & 40 & 59 &  1.46$\pm$0.0115 & 0.59$\pm$0.07 & 4 & 7 & 87 \\
HZ 43 B
 & 44 & 55 &  1.46$\pm$0.0101 & 0.66$\pm$0.07 & 5 & 6 & 87 \\
\textbf{Average} & 49 & 50 &  1.46$\pm$0.0093 & 0.61$\pm$0.11 & 3 & 3 & 93 \\
\hline
14 Aur Cb & 38 & 61 &  1.35$\pm$0.0193 & 0.49$\pm$0.13 & 3 & 6 & 89 \\
14 Aur Cb & 33 & 66 &  1.33$\pm$0.0225 & 0.60$\pm$0.18 & 3 & 6 & 89 \\
\textbf{Average} & 55 & 44 &  1.34$\pm$0.0131 & 0.54$\pm$0.09 & 6 & 4 & 88 \\
\hline
HD 2133 B
 & 25 & 74 &  1.42$\pm$0.0157 & 0.49$\pm$0.09 & 2 & 8 & 88 \\
HD 2133 B
 & 22 & 77 &  1.41$\pm$0.0179 & 0.32$\pm$0.07 & 2 & 8 & 89 \\
\textbf{Average} & 39 & 60 &  1.42$\pm$0.0092 & 0.40$\pm$0.14 & 1 & 2 & 96 \\
\hline
Sirius B
 & 59 & 40 &  0.80$\pm$0.0036 & 0.96$\pm$0.03 & 14 & 9 & 76 \\
Sirius B
 & 58 & 41 &  0.80$\pm$0.0036 & 0.88$\pm$0.03 & 14 & 9 & 76 \\
Sirius B
 & 58 & 41 &  0.80$\pm$0.0036 & 0.92$\pm$0.03 & 13 & 9 & 76 \\
Sirius B
 & 59 & 40 &  0.81$\pm$0.0036 & 0.96$\pm$0.03 & 14 & 9 & 75 \\
\textbf{Average} & 21 & 78 &  0.80$\pm$0.0112 & 0.93$\pm$0.10 & 4 & 15 & 79 \\
\hline
HR 1358 B
 & 28 & 71 &  1.22$\pm$0.0123 & 0.75$\pm$0.07 & 4 & 12 & 82 \\
HR 1358 B
 & 28 & 71 &  1.25$\pm$0.0123 & 0.71$\pm$0.14 & 2 & 6 & 90 \\
\textbf{Average} & 20 & 79 &  1.24$\pm$0.0180 & 0.73$\pm$0.05 & 6 & 24 & 69 \\
\hline

\end{tabular}
\end{minipage}
\end{table*}


\subsection{Testing the MRR}

The combined results from both the Lyman and Balmer spectra support the validity of the MRR. In comparison to studies using less accurate parallaxes (e.g., \citealt{Schmidt96}), the scatter in the data and the errors on the data points are considerably reduced. Fig. \ref{fig:Hip_vs_gaia} shows that all but one of the stars which had a \textit{Hipparcos} parallax available previously have converged towards the MRR rather than staying in the same position but with smaller errors.
In Fig. \ref{fig:MRR}, most of the targets cluster around the 0.6 M$_{\odot}$ range. This is expected due to the sharply peaked WD mass distribution at $\sim0.6$ M$_{\odot}$. There is a lack of data at the high mass range except for Sirius B discussed in section \ref{section:Sirius_B}. The mass of HD 2133 B is uncertain as the \textit{HST} results are within 2$\sigma$ of the MRR but the \textit{FUSE} result is not consistent at the $>3\sigma$ level.

A detailed comparison of each white dwarf with a theoretical relation of the appropriate temperature reveals that only five out of eleven are within 2$\sigma$. This might indicate that the details of the MRR still need adjustment. However, it has been shown in section \ref{section:sources_of_uncertainty} that the uncertainty in individual spectra may be underestimated. Four out of the five which agree with the MRR have errorbars calculated from the spread in results from several spectra. Of the six outliers, five have errors based on the statistical uncertainty in a single spectrum. The errors were recalculated by assuming a log $g$ uncertainty of 0.1 which was found to be the spread in log $g$ from multiple spectra of HZ 43 B in Fig. \ref{fig:logg_spread}. With these more realistic errors, 45 per cent of the sample agree within 1$\sigma$ and 91 per cent are within 2$\sigma$. Although still slightly lower percentages than would be expected if the MRR were valid, this does not raise serious doubts about the validity of the MRR, given that there are still some doubts over the exact mass-radius values for HD 2133 B. We are currently analysing further data using the gravitational redshift method which may help to better constrain the mass of HD 2133 B. What the data does show, is that the uncertainties in most cases are now small enough to clearly identify WDs which follow the MRR, and also to distinguish stars which do not follow the MRR such as HD 15638.

For the 10 stars within 2$\sigma$ of the MRR, 3 favour a thin H-layer model. 14 Aur Cb is within 2$\sigma$ of the thin MRR and more than 3$\sigma$ from the thick MRR for both the UV and optical results. HZ 43 B and REJ 1925 are less certain as they are both within 1$\sigma$ of the thin model but still agree with the thick model within 2$\sigma$. HD 2133 B is closer to the thin MRR but with relatively large uncertainty as previously discussed. Of the remaining 7 stars, Feige 24 and HD 223816 are the only ones which are clearly a better fit to the thick H-layer MRR and are $>3\sigma$ from the thin MRR. The rest, including Sirius B, are within 2$\sigma$ of both thick and thin H-layer models and can not distinguish between them, although the dynamical mass of Sirius B (see section \ref{section:Sirius_B}) clearly agrees with the thick H-layer MRR. This adds to the findings of previous studies by \cite{Provencal98} and \cite{Romero_2012} which have also found evidence for a range of H-layer thickness for DA white dwarfs.


\begin{figure}
\includegraphics[width=84mm]{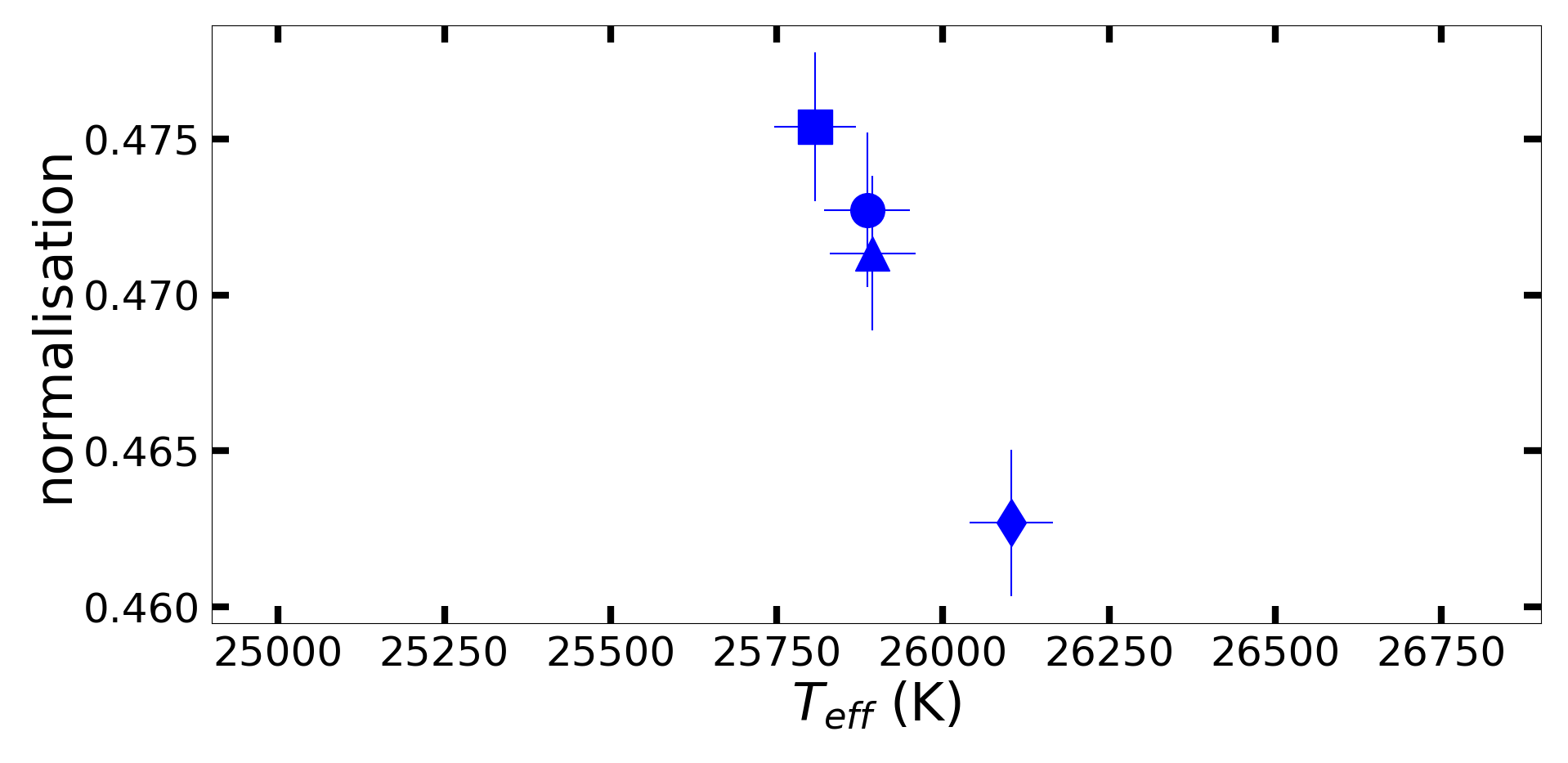}
\includegraphics[width=85mm]{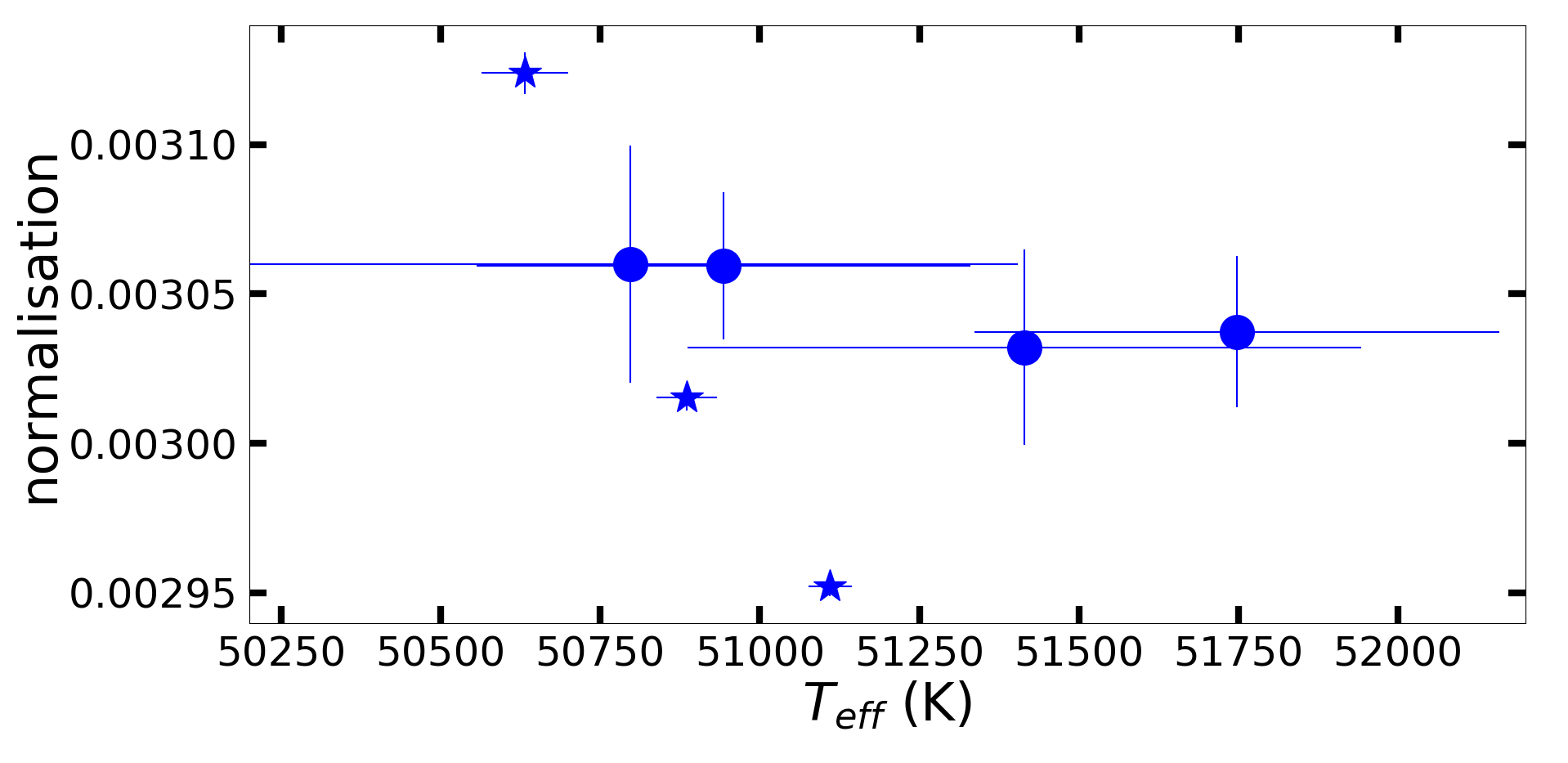}

\caption{\textbf{Top panel:} The spread in normalisation from fitting 4 \textit{HST} spectra of Sirius B. The results from spectrum obt801010 (diamond) gives a slightly higher temperature than the other 3 spectra.
\textbf{Bottom panel:} The spread in normalisation from fitting 4 \textit{HST} (circles) and 3 \textit{FUSE} spectra (stars) of HZ 43 B. Results are consistent between the Lyman and Balmer spectra.}
\label{fig:norm_spread}
\end{figure}

\begin{figure}
\includegraphics[width=83.5mm]{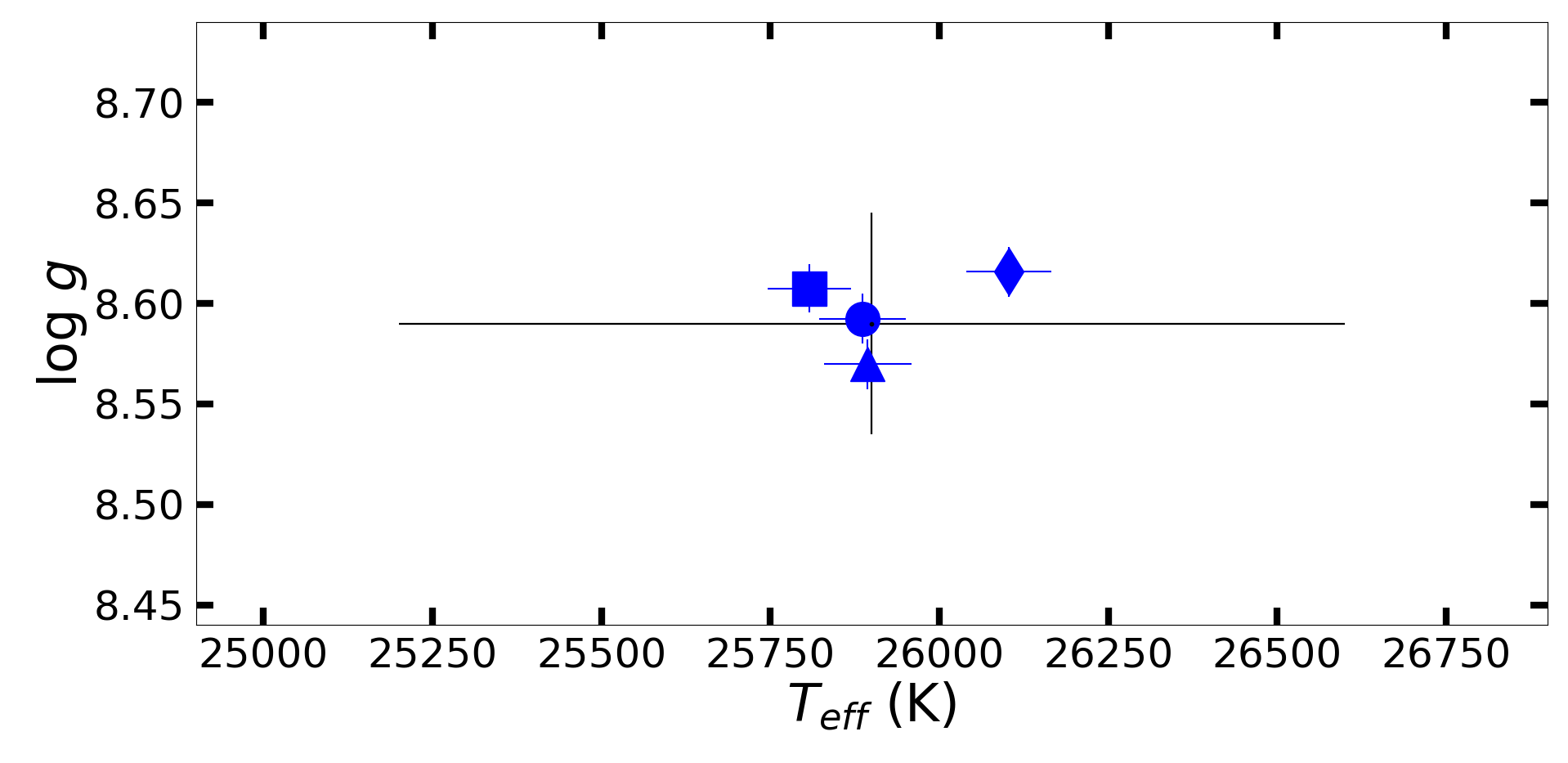}
\includegraphics[width=84mm]{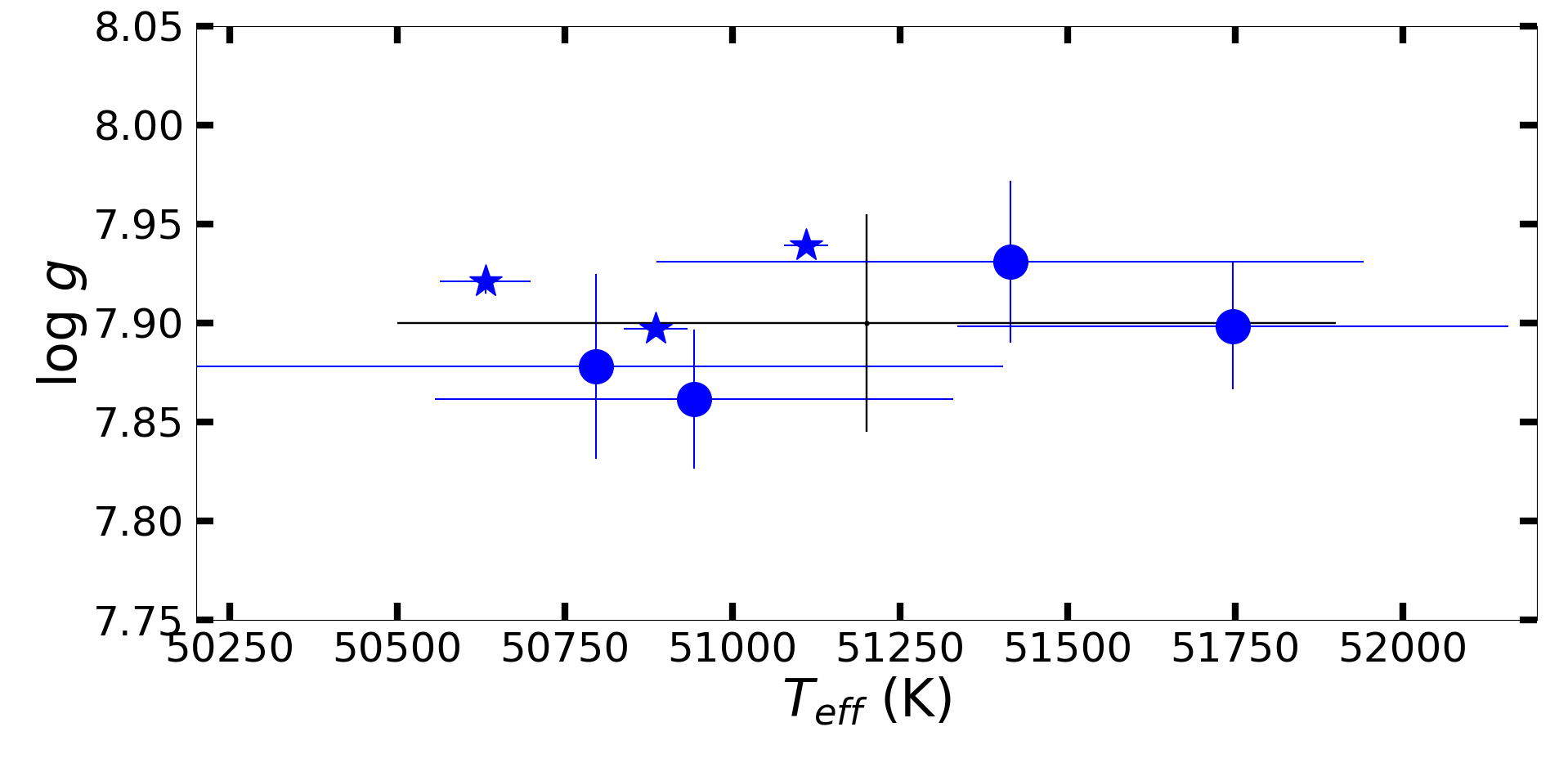}

\caption{Scatter in log $g$ and $T_{\rm eff}$ parameters measured from multiple spectra of the same star. For comparison, the black cross in both panels indicates the spread found for WD1327$-$083 (Wolf 485A) as shown in Fig. 6 of Tremblay el al. (2017). 
\textbf{Top panel:} Sirius B, best fit parameters for 4 \textit{HST} Balmer line spectra. 
\textbf{Bottom panel:} HZ 43 B, \textit{FUSE} Lyman line spectra (stars) and \textit{HST} Balmer line spectra (circles).}
\label{fig:logg_spread}
\end{figure}

\begin{figure}
\includegraphics[width=84mm]{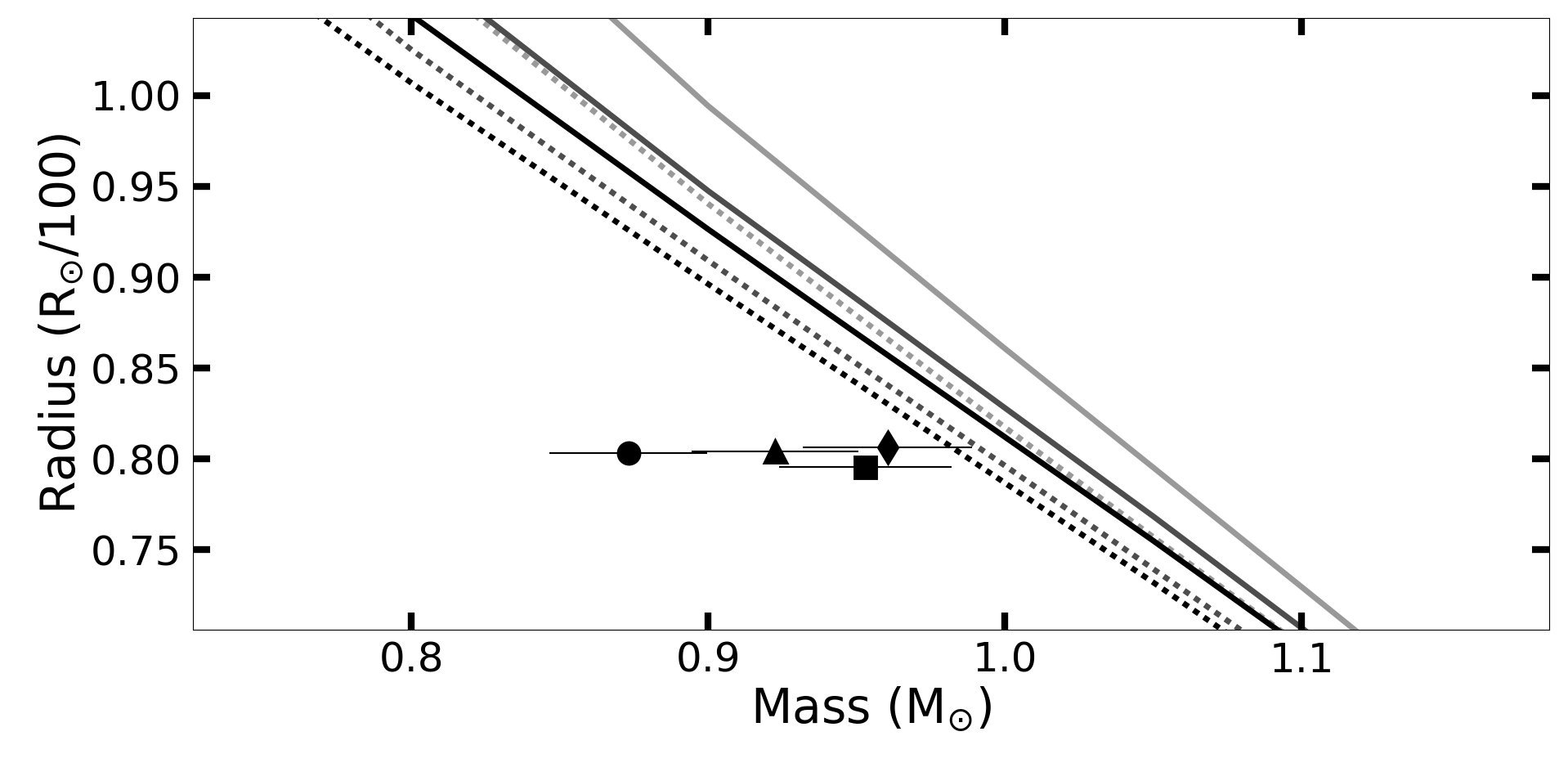}
\includegraphics[width=84mm]{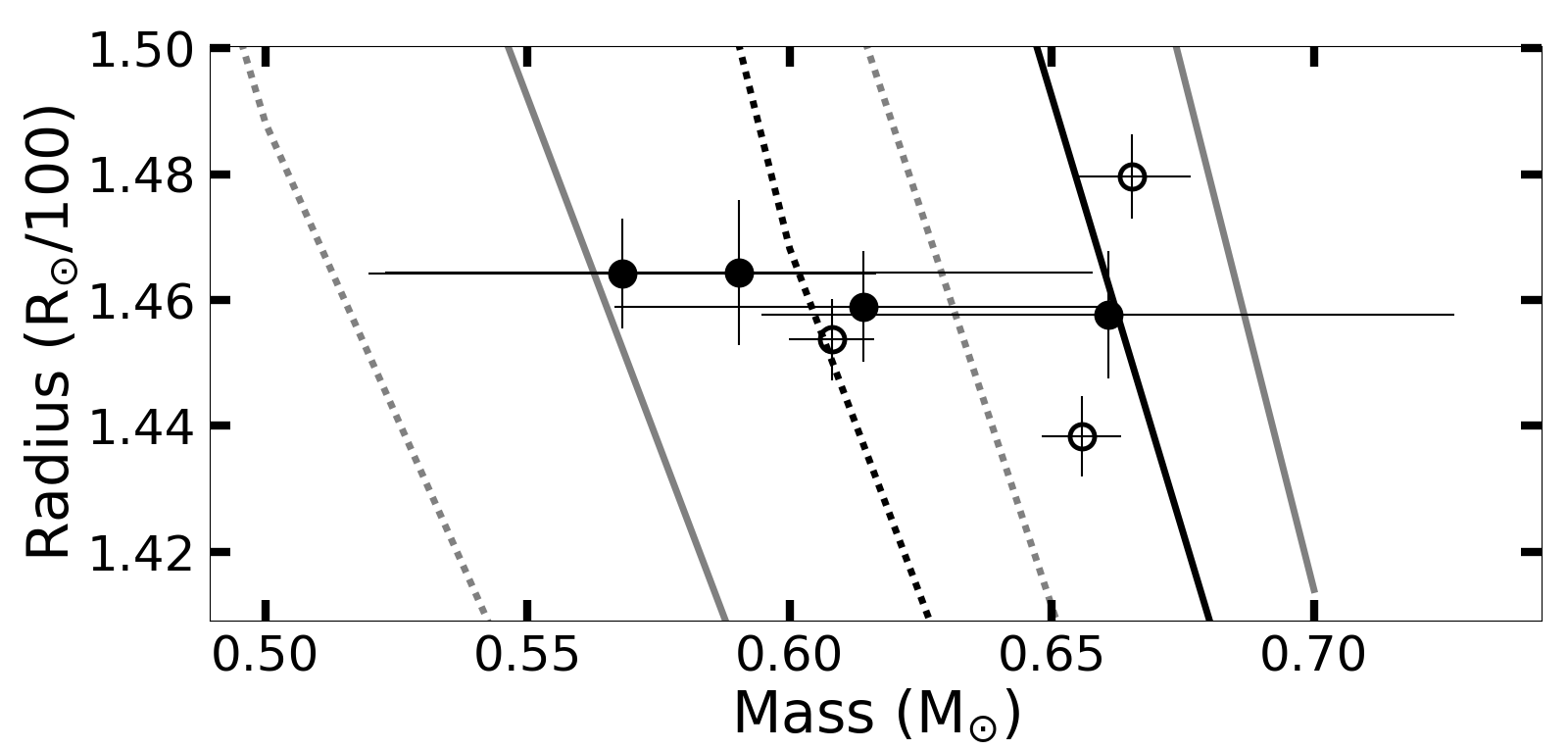}

\caption{\textbf{Top panel:} The MRR results for each spectrum of Sirius B calculated individually. All results are from \textit{HST} Balmer line spectra.
\textbf{Bottom panel:} Results for each spectrum of HZ 43 B including Lyman line (empty circles) and Balmer line spectra (filled circles). The $T_{\rm eff}$ measured for HZ 43 B is 51,189 K. The MRR is plotted for temperatures of 25,000, 51,000 and 58,000 K from left to right. The data points are clustered around the MRR for 51,000 K but are consistent with both thin (dashed line) and thick (solid) models.}
\label{fig:MRR_spread_Sir_HZ 43}
\end{figure}



\subsubsection{The effect of temperature}

According to the theoretical models, some spread in the mass-radius results is expected due to the different temperatures of the stars in the sample. It is expected that WDs of a given mass will have a larger radius if they have a higher temperature. The limited size of the sample means there are not enough stars in each temperature range to test the MRR across the full mass range. However, we can test for any general trends with temperature using the full sample.

In Fig. \ref{fig:MRR_Teff} we examine the question of whether the data are better represented by the zero temperature MRR or by models which include the effect of temperature. Fig. \ref{fig:MRR_Teff} shows the result of plotting the MRR with markers colour coded according to 5 temperature bins. They are compared to theoretical MRR tracks for a range of temperatures and thick hydrogen layers. The results match the expected trend with the lowest temperature stars slightly below the zero temperature relation and hotter stars increasingly further above the zero temperature relation.
The results for Feige 24 and HD 223816 are the most inconsistent with the zero temperature MRR. These two stars are also the hottest in this sample with temperatures of 62,835 K and 73,400 K respectively. 

The correlation between increasing temperature and radii larger than the zero temperature model indicates that temperature is indeed an important factor. Much better agreement is found when these targets are compared to the MRR appropriate for their temperature. Only three are consistent with the zero-temperature carbon core model compared to ten when temperature effects are included.
Although the data matches the expected trend with increasing temperature, it appears that there is a wider spread in possible radii for a given mass than would be expected from the models, even when the thick H-layer models, which give larger radii, are used. For example, Feige 24 lies above the MRR while 14 Aur Cb and HZ 43 B lie below the MRR for their temperature despite all 3 having almost the same mass. They would be expected to have a similar core composition given their similar mass. This may indicate that the influence of temperature on the radius is slightly underestimated in current models. 


\subsubsection{Comparison of Lyman and Balmer line results}

\begin{figure*}
\includegraphics[width=175mm]{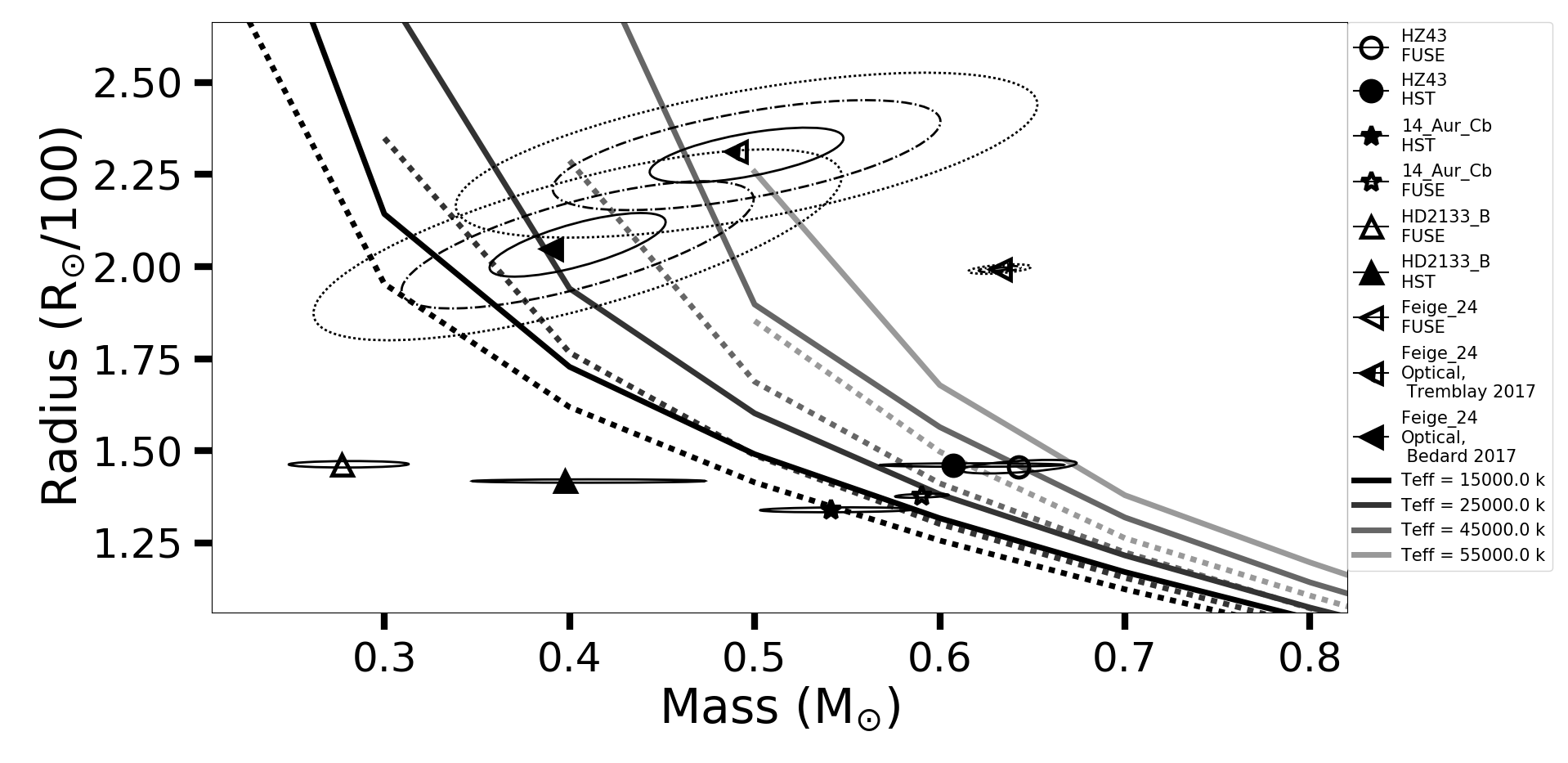}
 
 \caption{Comparison of mass-radius results for 4 WDs with both far-UV (hollow markers) and optical (solid) spectra. The error ellipses are 1$\sigma$ for most targets except for Feige 24 where 1$\sigma$, 2$\sigma$ and 3$\sigma$ ellipses are shown. The size of the ellipses for the Feige 24   optical data points are based on the log $g$ and radius (with DR1 parallax) uncertainties quoted in \citep{Tremblay_eta17, Bedard17}.}
  \label{fig:Lyman_vs_Balmer}
\end{figure*}

A further issue to examine is the possibility of systematic differences between values derived from the Lyman or Balmer spectra. In the analysis by \cite{Barstow_eta03}, good consistency was found when comparing Lyman and Balmer results for WDs below 50,000 K  using the H/He grid based on the Lemke broadening tables, but increasing discrepancies in temperature became apparent above this temperature. Here we repeat this test using the new \cite{Tremblay09} model grid.
There are 4 WDs where we can compare the Lyman and Balmer results covering a T$_{\rm eff}$ range of 29,700 K to 62,835 K. HZ 43 B, 14 Aur Cb and HD 2133 B were observed by both \textit{FUSE} and \textit{HST}. For Feige 24 we include the Balmer line results obtained by \cite{Tremblay_eta17} and \cite{Bedard17} to compare to our \textit{FUSE} data.

Fig. \ref{fig:Lyman_vs_Balmer} compares the mass-radius values obtained for each target. Each target is plotted twice with solid markers for optical data and hollow markers for far-UV.  The uncertainty is represented as 1$\sigma$ error ellipses which are at an angle to the axis due to the $R^{2}$ being included in the mass equation. 

For the WDs below 50,000 K we found no significant difference in the results obtained from the same target when using both the Lyman and Balmer lines except in the case of HD 2133 B. 
HZ 43 B and 14 Aur Cb both have Balmer and Lyman line results in agreement within the errors. HZ 43 B shows that the results from Balmer and Lyman line fitting, even at 50,000 K,  are entirely consistent (plotted as circles in Fig. \ref{fig:Lyman_vs_Balmer}). A more detailed plot showing the results for all the HZ 43 B spectra individually is shown in Fig. \ref{fig:norm_spread} and \ref{fig:logg_spread}, (lower panel). The 3 Lyman line spectra (stars) and 4 Balmer line spectra (circles) are in excellent agreement despite coming from different instruments and different wavelength ranges. 

Feige 24 is above 50,000 K which was noted as the boundary where Lyman and Balmer determinations start to diverge \citep{Barstow_eta03}. Both our far-UV and the optical result of \cite{Tremblay_eta17} lie near the MRR for the temperature of 55,000 K as expected. However, there is no agreement between any of the 3 optical and UV data points at $>3\sigma$.

At such high temperatures, radiative levitation can bring up trace amounts of heavy metals which could affect the shape of the hydrogen absorption lines. This would alter the log $g$ and $T_{\rm eff}$ derived from their fitting. 
Feige 24 was one of the targets studied by \cite{Barstow_eta14} and found to contain heavy metals.

\cite{Barstow_eta14} using models including heavy metals but with the older Lemke broadening tables found a lower log $g$ of 7.53. Combined with the new distance measurement, this gives a mass of 0.46 M$_{\odot}$ which is still not compatible with the optical data and no longer lies on the MRR at 55,000 K. More work is needed to understand the effects of heavy metals on the spectroscopic parameters and to solve the Lyman-Balmer problem \citep{Preval_2015} before WDs above 50,000 K can reliably constrain the MRR.


\subsubsection{Details of individual systems}\label{section:individual_systems}
Several of the targets in this sample had very uncertain distance information available previously. 14 Aur Cb, HD 2133 B and HR 1358 B are also unresolved from the ground and these are the first optical spectra to be obtained. We review what the new data tell us about these WDs and the implications for the MRR here.

\textbf{HD 15638: }\label{section:HD 15638} The DR1 parallax was 3.8 $\pm$ 0.3 mas making this star a notable $>3\sigma$ outlier in both mass and radius. The DR2 parallax is almoust double the previous value at 6.4 $\pm$ 0.2 mas and has moved this data point from the top right of Fig. \ref{fig:MRR} to the bottom left. The \textit{FUSE} spectrum was previously analysed by \cite{Kawka_2010} who found log $g$ and $T_{\rm eff}$ values in agreement with ours. Their mass estimate of 0.54 $\pm$ 0.01 M$_{\odot}$ was derived using the MRR because no distance measurement was available. With the new \textit{Gaia} parallax the star is more than 2$\sigma$ below the C/O core MRR.

\textbf{14 Aur Cb:} The results for 14 Aur Cb show that it is close to the average WD mass. The Balmer and Lyman spectra give a mass of 0.54 $\pm$ 0.09 and 0.59 $\pm$ 0.03 M$_{\odot}$ respectively. Comparison with the MRR for a C/O WD at 45,800 K shows that the data matches the thin H-layer models within 1$\sigma$ and is not consistent with the thick H-layer model. 

\textbf{HR 1358 B:} was discovered by \cite{Boehm93} and is a member of the Hyades cluster. Previous analysis \citep{Burleigh_eta98} found a mass of 0.98 M$_{\odot}$ which was noted as probably too high because it gave a total age for the system which was younger than the cluster age.  From the \textit{HST} spectra we find a lower mass of 0.73 $\pm$ 0.05 M$_{\odot}$.

A lower mass also increases the estimated age of the system due to the longer main sequence lifetime of a lower mass WD progenitor. For a WD mass of 0.73 M$_{\odot}$ and $T_{\rm eff}$ 20,900 K, the total main sequence plus WD cooling age is approximately 240 Myrs. This is still some way short of the Hyades cluster age of 625 $\pm$ 50 Myr \citep{Perryman_eta98}. Even with the lower mass estimate of 0.73 M$_{\odot}$, HR 1358 B is still more massive than the majority of WDs, which makes this an important target for constraining the sparsely sampled high-mass end of the MRR. 

\textbf{HD 2133 B:} \label{section:HD 2133}
The mass-radius results for HD 2133 B are still somewhat uncertain. It was noted that the results for the \textit{FUSE} spectrum of HD 2133 B changed from 0.61 M$_{\odot}$ when fitting with models based on the older Lemke broadening tables, down to only 0.28 M$_{\odot}$ with the newer Tremblay based models. However the change for the \textit{HST} spectra was much smaller, from 0.43 down to 0.4 M$_{\odot}$. The \textit{FUSE} mass of 0.27 M$_{\odot}$ is lower than would be expected from single star evolution, which implies that the \textit{HST} mass is more likely to be correct. It should be noted that 0.4 M$_{\odot}$ is the average of the 2 \textit{HST} spectra and there is a range of 0.2 M$_{\odot}$ between them. With this large uncertainty, HD 2133 B may be considered consistent with the C/O core MRR. Previous far-UV results \citep{Burleigh_eta97} support the higher mass estimate (0.6 $\pm$ 0.05 M$_{\odot}$). 

\textbf{RE 0357:} With the DR1 parallax this WD fell significantly below the MRR. DR2 increased the distance derived from the parallax from 98 to 108 pc and has brought this result into agreement with the MRR. This system includes a K2V main sequence star which is known to be very rapidly rotating despite being old enough to have spun down. The scenario suggested to explain this is that the MS star accreted material during the AGB phase of the current WD companion. This system was first noted as a UV excess source by \cite{Jeffries_eta96}. The UV emission of a hidden WD was suggested as the true origin of the excess UV luminosity of the MS star. The required WD parameters in this scenario were calculated by \cite{Jeffries_eta96} as mass $=$ 0.4$-$0.7 M$_{\odot}$ and $T_{\rm eff} =$ 30,000$-$40,000 K. Our \textit{FUSE} results fall within the predicted range with mass = 0.54 $\pm$ 0.04 M$_{\odot}$ and $T_{\rm eff}$ = 33,927 $\pm$ 66 K.

A higher mass of 0.79 M$_{\odot}$ was found from the EUV results \citep{Burleigh_eta97} using the \textit{Gaia} distance of 108 pc, although that requires a log $g$ of 8.25 which is incompatible with the log $g$ of 7.87 found here and by \cite{Barstow_eta14}.


\subsection{Sirius B}\label{section:Sirius_B}

\begin{table*}
\begin{minipage}{140mm}

\caption{Comparison of mass and radius results for Sirius B from spectra analysed by various authors and the mass derived from astrometric measurements.}
\label{table:Sirius_results}
\begin{tabular}{lcc}

\hline

Method/model & Radius  & Mass \\
&(0.01 R$_{\odot}$) & (M$_{\odot}$) \\
\hline

Dynamical mass, \citep{Bond17} & - & 1.018 $\pm$ 0.011\\

Spectroscopic, Tremblay model (this paper) & 0.80 $\pm$ 0.01 & 0.927 $\pm$ 0.107\\

Spectroscopic, HST (2004 data, \citealt{Barstow05}) &0.80 $\pm$ 0.04&0.841 $\pm$ 0.08\\
Spectroscopic, HST (2004 data, \citealt{Bedard17}) &0.79 $\pm$ 0.02&0.940 $\pm$ 0.11\\
Spectroscopic, Ground based \citep{Tremblay_eta17}\footnote{Spectroscopic parameters derived from the ground based spectrum described in \cite{Gianninas_2011}.} & 0.80 $\pm$ 0.01 & 0.872 $\pm$ 0.084\\

\hline
\end{tabular}
\end{minipage}
\end{table*}

\begin{figure}
\includegraphics[width=83mm]{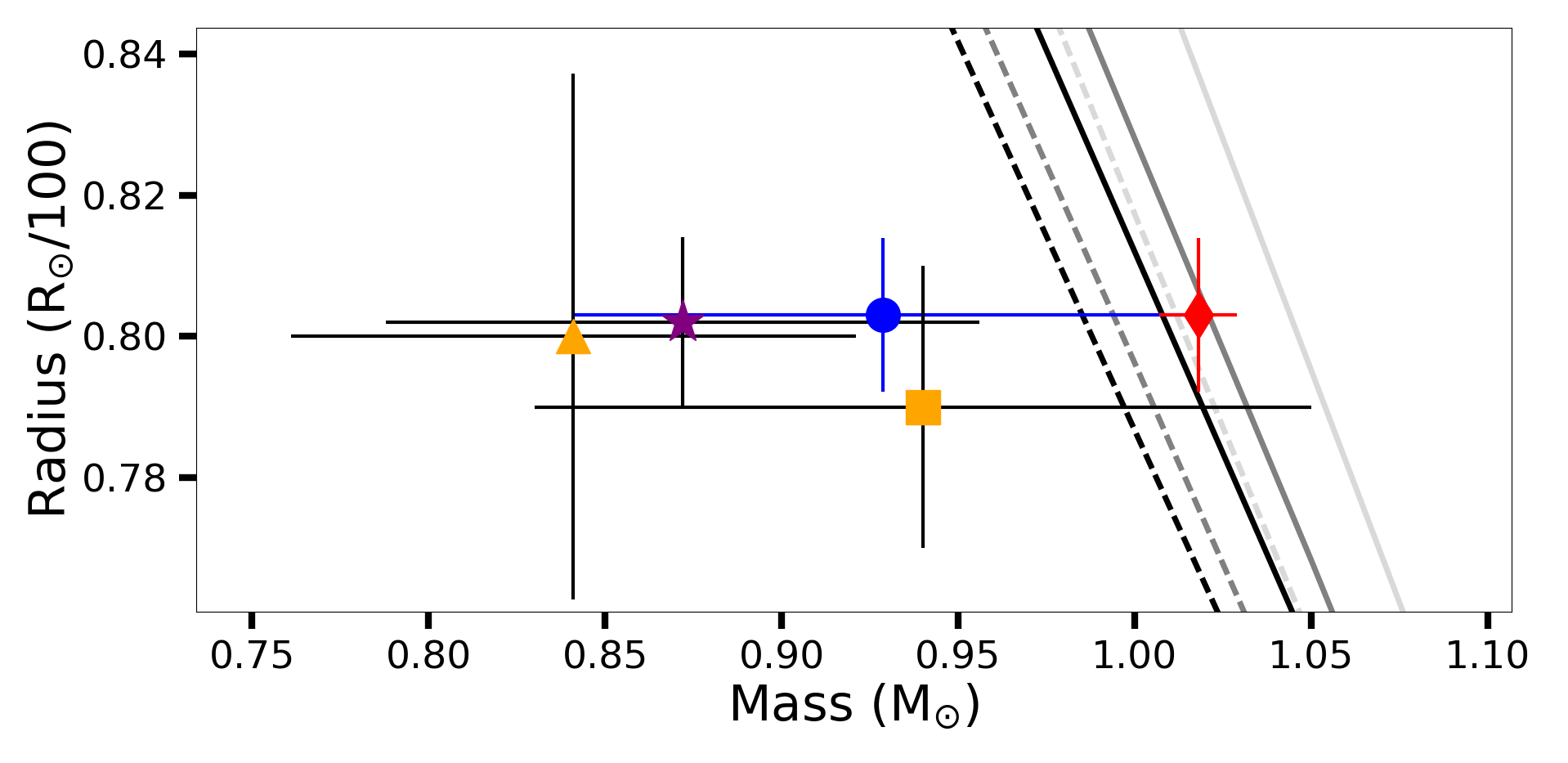}

\caption{Sirius B. Comparison of mass measured using the dynamical (Red diamond) and spectroscopic method. The dynamical mass (Bond et al 2017) is consistent with the MRR for a C/O core WD at 25000 K. The theoretical tracks are from darkest to lightest 15,000, 25,000 and 45,000 K. Dashed lines are thin H envelope and solid lines are thick H envelope. The yellow markers are the 2005 \textit{HST} Balmer line spectrum. The B\'edard et al. (2017) result (yellow square) is consistent with the results from the new spectra (Blue circle) and give a larger mass than the Barstow et al. (2005) result using the same spectrum (yellow triangle). Also plotted is the Tremblay et al. (2017) measurement from a ground based spectrum (purple star).}
\label{fig:Sir_comp_methods}
\end{figure}


Sirius B represents an important benchmark for validating the results of the spectroscopic mass measurements because its mass can be measured by several independent methods. The mass is most accurately determined from the dynamical method which uses long term observations of the binary orbit to calculate the complete set of binary parameters including the mass of both Sirius A and B. \cite{Bond17} recently published the results of analysis of 150 years of observations of the Sirius system, including almost 20 years of \textit{HST} observations. From these it has been possible to measure the orbital motion of Sirius B over most of its 50 year orbit. 
The mass calculated in this way is the most reliable because it depends only on well-known laws of mechanics rather than spectral modelling. The disadvantage of this method is that it doesn't provide any information on the radius. 

In Fig. \ref{fig:Sir_comp_methods} we have combined the \cite{Bond17} dynamical mass (1.018 $\pm$ 0.011 M$_{\odot}$) with the radius measured from our spectral fitting (red diamond). The radius measurement is dependent on the normalization from spectral fitting and the distance from the parallax which are two of the most accurately measured parameters in our method. It should be noted that there is no parallax for Sirius in \textit{Gaia} DR2 because the star is too bright to be handled by the normal processing. The parallax used here is the \textit{Hipparcos} value $379.21 \pm 1.58$ mas from the new reduction of the catalogue \citep{vanLeeuwen07}. The uncertainty on this value is already very small and it is not expected to change much when measured by \textit{Gaia}. The fitting of the 4 \textit{HST} spectra for Sirius B showed very little variation in the radius and so we were able to calculate an average radius value of comparable accuracy to the dynamical mass. Table \ref{table:mass_radius} and \ref{table:Sirius_results} list the calculated mass and radius values with associated errors. 

Comparison with the theoretical MRR reveals that the dynamical mass with the spectroscopic radius is in excellent agreement with the MRR for a C/O core WD with a temperature of 25,900 K and a thick hydrogen layer. This is one of the few measurements where the uncertainty is small enough to be able to show clearly that the data is consistent with a thick H-layer model and not with the equivalent thin H-layer model of the same temperature. 
The mass measured using the spectroscopic method is 0.93 $\pm$ 0.1 M$_{\odot}$ which is almost 10 per cent less than the dynamical mass. The two measurements are consistent within 1$\sigma$, although this is due to the relatively large uncertainty in the spectroscopic mass resulting from the spread in log $g$ values from the 4 spectra. By itself, this result does not indicate any serious discrepancy between the two methods. However, a comparison with the results found by other studies \citep{Tremblay_eta17, Bedard17} shows that the spectroscopic mass is consistently lower by at least 10 percent.

It is possible that improvements to the models may increase the mass estimate. Evidence for systematic differences being dependent on the models used rather than random error is found from comparing the results obtained by two different studies which both used the same \textit{HST} spectrum of Sirius B taken in 2004. This spectrum was fitted by \cite{Barstow05} and has recently been re-analysed by \cite{Bedard17}. The models used are both pure hydrogen NLTE. The main difference between the studies is that \cite{Barstow05} used models based on the \cite{Lemke97} broadening tables, whereas \cite{Bedard17} used the \cite{Tremblay09} tables and included 3D corrections \citep{Tremblay13}. It can be seen in Fig. \ref{fig:Sir_comp_methods} that the 2005 analysis resulted in a spectroscopic mass of only $0.841 \pm 0.08$ M$_{\odot}$ which is incompatible with the dynamical mass. The 2017 analysis has resulted in an increased mass of $0.94\pm0.11$ M$_{\odot}$ which is still lower than the dynamical mass, but is in close agreement with the results from the more recent \textit{HST} spectra presented in this paper. It is interesting that the spectroscopic results from 2 different sets of \textit{HST} spectra analysed independently, as well as the analysis by \cite{Tremblay_eta17} using the ground based spectrum from \cite{Gianninas_2011}, give results in complete agreement with each other, but consistently lower than the dynamical mass by $0.08$ M$_{\odot}$. The results are also in disagreement with the mass derived from the gravitational red-shift method \citep{Barstow_Joyce2017} which is 1.1 $\pm$ 0.03 M$_{\odot}$. 

It can be concluded that, although spectroscopic masses are formally consistent with the dynamic mass at the 1$\sigma$ level, they are systematically about 10 per cent lower than the dynamic mass. It is more likely that the models, rather than the data or fitting method, are responsible for the systematic offset. The models provide a good fit to the data, but consistently give a gravity which is too low, causing the mass to be underestimated. It is hoped that these results from Sirius B might also help to indicate where improvements to the models need to be made. 

One of the keys to resolving this issue may be the use of lab based tests to validate the theoretical models. Falcon et al. (2017) have developed laboratory tests which can probe higher plasma densities and have shown that even for the updated broadening tables \citep{Tremblay09}  there are still differences between the theoretical and observed line profiles when the density of the plasma is increased to the levels found in WD atmospheres. Improved treatment of the Stark effects, which have been shown to cause asymmetry in the Balmer lines \citep{Halenka2015} may be necessary to correctly fit the high quality spectra now available.


\section{Conclusions}
We have conducted a study with the main aims of testing the MRR using state of the art data and comparing results from optical and far-UV spectroscopy. The detailed analysis of the uncertainties involved with the Balmer/Lyman line fitting are a step towards improving upon the results that can currently be achieved. In particular, the validation of the Lyman line results will make it possible to extend spectroscopic studies to many SLSs which cannot be studied in the optical.

The use of parallax data from \textit{Gaia} DR2 has substantially reduced the uncertainty in the mass determinations and also enabled us to obtain new results for some systems which previously had no parallax measurement available. 

In common with studies using DR1 (\citealt{Tremblay_eta17, Bedard17}), we find that most WDs in the sample are consistent with the MRR. 91 per cent of the WDs studied are within 2$\sigma$ of the theoretical MRR appropriate for their temperature when realistic uncertainties are considered. HD 15638 is the main WD which does not agree within 2$\sigma$ and is noted as having a significantly different parallax in DR1 compared to DR2. As shown by \cite{Bedard17}, the improvement in precision makes it possible to identify individual WDs which are inconsistent with the general trend of the MRR followed by most of the sample. 2 stars in the sample are a better fit to thin H-layer models while others agree with thick H-layer models. This is similar to the findings of \cite{Provencal98} and \cite{Romero_2012} and shows that a range of evolutionary scenarios may have to be considered.

We have also confirmed the finding of \cite{Barstow_eta03} that Lyman line fitting produces results consistent with the Balmer line fitting within 1$\sigma$. However, this does not apply to WDs above 50,000 K. More work is needed to understand the cause of the divergence in spectroscopic parameters obtained for WDs above 50,000 K, particularly the role that trace heavy metals may play in altering the shape of the hydrogen lines.

Despite using the best available space based spectra, this study agrees with the conclusions of \cite{Tremblay_eta17} which showed that spectroscopic tests of the MRR are now limited by the accuracy of the spectroscopically derived parameters rather than the parallax. The spread in results obtained from multiple spectra of Sirius B and HZ 43 B highlight the fact that even with the best optical spectra, the uncertainty still makes definitive tests of the MRR problematic. For HZ 43 B which has 7 spectra available, there is a spread of 0.1 in log $g$ which is similar to that found for Wolf 485A \citep{Tremblay_eta17}. As a consequence, there is an uncertainty in the mass derived from spectral fitting of at least 0.1 M$_{\odot}$ even for high S/N spectra. This is larger than the statistical uncertainty found from fitting single spectra by a factor of $\sim$2-3. The log $g$ parameter is the main contributor. From the spread in HZ 43 B results, it is estimated that the mass can currently only be measured to a precision of $\sim 14$ per cent using Balmer line and 10 per cent using Lyman line data without using the MRR to derive the mass. 

A preliminary attempt to search for observational evidence of the predicted temperature dependence of the WD radius in the data was not conclusive. The temperature clearly has some effect, with the hottest stars having the largest radii for a given mass. However, a much larger sample covering a wide temperature range for each mass bin will be required to provide a detailed test. 

The radius obtained for Sirius B has produced a result in firm support of the MRR at the high mass end when combined with the dynamical mass (1.018 $\pm$ 0.011 M$_{\odot}$) of \cite{Bond17}. The spectroscopic mass is formally within 1$\sigma$ of the dynamical mass and the MRR, although there is an apparent tendency for the mass to be underestimated, and the spread in spectroscopic results is much larger than the uncertainty in the dynamical mass. 

In order to make progress with testing the MRR, the following issues will need to be addressed. Firstly, the causes of uncertainty in parameters derived from hydrogen line fitting will need to be identified and reduced. This includes the spread in results from different observations of the same target, which will most likely require improved spectra and refinements to the methods of fitting them. It also includes a more wide-ranging investigation into possible systematic offsets, especially for high mass white dwarfs and those above 50,000 K. Secondly, it will be necessary to extend the comparison of mass estimates obtained from different methods to include many more systems. To this end we are carrying out gravitational redshift studies of several SLSs including Sirius B. Dynamical studies will continue to be an important bench mark, although orbital determinations for more systems will take time to accumulate. \textit{Gaia} will potentially help in this endeavour by providing high precision measurements which can constrain a long period orbit in a much shorter time than has previously been possible.

\section*{Acknowledgements}

We would like to thank the referee for helping us to make significant  improvements to the text. SRGJ acknowledges support from the Science and Technology Facilities Council (STFC, UK). 
MAB acknowledges support from the Gaia post-launch support programme of the UK Space Agency.
This work has made use of data from the European Space Agency (ESA)
mission {\it Gaia}, processed by
the {\it Gaia} Data Processing and Analysis Consortium (DPAC). Funding
for the DPAC has been provided by national institutions, in particular
the institutions participating in the {\it Gaia} Multilateral Agreement.
JBH was partially supported by NSF grant AST-1413537

\appendix

\bsp

\label{lastpage}

\end{document}